\documentclass[]{aastex631}

\DeclareRobustCommand{\okina}{%
  \raisebox{\dimexpr\fontcharht\font`A-\height}{%
    \scalebox{0.8}{`}%
  }%
}
\newcommand{\Ou}{{\okina}Oumuamua}

\usepackage{amsmath}
\newcommand{\kms}{$\mathrm{km}\ \mathrm{s}^{-1}$}

\def\sigdist#1#2{$#1\ \mathrm{km}\ \mathrm{s}^{-1} \pm#2\ \mathrm{dex}$}
\def\cl#1#2#3{\begin{tabular}{l}$M_\mathrm{min}=#1$\\ $M_\mathrm{max}=#2$\\ $\beta=#3$\end{tabular}}
\def\heat#1#2{\begin{tabular}{l}$\mathcal{H}=#1$\\ $\alpha_H=#2$\end{tabular}}

\shorttitle{ISO Streams}
\shortauthors{Forbes et al.}

\graphicspath{{./}{figures/}}

\begin{document}

\title{He awa whiria: the tidal streams of interstellar objects}

\author[0000-0002-1975-4449]{John C. Forbes}
\affil{School of Physical and Chemical Sciences--Te Kura Mat\=u, University of Canterbury, Christchurch 8140, New Zealand}

\author[0000-0003-3257-4490]{Michele T. Bannister}
\affil{School of Physical and Chemical Sciences--Te Kura Mat\=u, University of Canterbury, Christchurch 8140, New Zealand}

\author[0000-0001-5578-359X]{Chris Lintott}
\affiliation{Department of Physics, University of Oxford, Denys Wilkinson Building, Keble Road, Oxford, OX1 3RH, UK}

\author[0009-0008-0355-5809]{Angus Forrest}
\affil{School of Physical and Chemical Sciences--Te Kura Mat\=u, University of Canterbury, Christchurch 8140, New Zealand}

\author[0000-0001-5839-0302]{Simon Portegies Zwart}
\affil{Sterrewacht Leiden, Leiden University, Einsteinweg 55, 2333CC Leiden, The Netherlands}

\author[0000-0002-8910-1021]{Rosemary C. Dorsey}
\affil{School of Physical and Chemical Sciences--Te Kura Mat\=u, University of Canterbury, Christchurch 8140, New Zealand}

\author[0009-0003-4251-2821]{Leah Albrow}
\affil{Department of Earth Atmospheric and Planetary Sciences, Massachusetts Institute of Technology, Cambridge, MA, USA}
\affil{School of Physical and Chemical Sciences--Te Kura Mat\=u, University of Canterbury, Christchurch 8140, New Zealand}

\author[0000-0001-6314-873X]{Matthew J. Hopkins}
\affiliation{Department of Physics, University of Oxford, Denys Wilkinson Building, Keble Road, Oxford, OX1 3RH, UK}
\affil{School of Physical and Chemical Sciences--Te Kura Mat\=u, University of Canterbury, Christchurch 8140, New Zealand}

\correspondingauthor{John C. Forbes}
\email{john.forbes@canterbury.ac.nz}

\begin{abstract}

Upcoming surveys are likely to discover a new sample of interstellar objects (ISOs) within the Solar System, but questions remain about the origin and distribution of this population within the Galaxy.
ISOs are ejected from their host systems with a range of velocities, spreading out into tidal streams --- analogous to the stellar streams routinely observed from the disruption of star clusters and dwarf galaxies. 
We create a simulation of ISO streams orbiting in the Galaxy, deriving a simple model for their density distribution over time. 
We then construct a population model to predict the properties of the streams in which the Sun is currently embedded. 
We find that the number of streams encountered by the Sun is quite large, $\sim 10^6$ or more. 
However, the wide range of stream properties means that for reasonable future samples of ISOs observed in the Solar System, we may see ISOs from the same star (``siblings''), and we are likely to see ISOs from the same star cluster (``cousins''). 
We also find that ISOs are typically not traceable to their parent star, though this may be possible for ISO siblings.
Any ISOs observed with a common origin will come from younger, dynamically colder streams. 

\end{abstract}

\section{Introduction} \label{sec:intro}

Interstellar objects (ISOs) are unbound from their host planetary system and then orbit in the Galactic potential.
A given point in the Galaxy will then encounter a flux of ISOs.
Within the observable volume of the Solar System, there is a high enough spatial density of such objects that they occasionally pass close enough for detection and characterization, with two known at present \citep[e.g.][]{meech_Brief_2017, fitzsimmons_Spectroscopy_2018, bannister_ColOSSOS_2017,guzik_Initial_2020}. 
Understanding the factors that sculpt the phase space density of the Galactic population is key for using the sensitivity of well-characterized sky surveys to place constraints on either the local or the general ISO number density \citep[e.g.][]{engelhardt_Observational_2017,jewitt_Interstellar_2017}. 
\citet{meech_Brief_2017} and \citet{do_Interstellar_2018} estimated a local density of $\sim 0.2\ \mathrm{au}^{-3}$, equivalent to $\sim 10^{16}$ objects produced per star in the Milky Way. 
At present, the estimates of the background density of ISOs are inherently uncertain at the order-of-magnitude level.
Both well-characterized surveys and improvements in the entirely reasonable modelling assumptions employed to date are necessary. 
Such high local densities may produce of order $\sim10^{2}$ discoveries of ISOs in the upcoming surveys by the Vera C. Rubin Observatory \citep{Ivezic_LSST_2019,schwamb_Large_2018a} and NEOSurveyor \citep{mainzer_Earth_2023}.

Several refinements can be made with the aim of making testable predictions and understanding the roles ISOs play in the Galaxy \citep{pfalzner_Hypothesis_2019,moro-martin_Could_2019,pfalzner_Oumuamuas_2020}, and what we can learn from them \citep{laughlin_Consequences_2017, portegieszwart_Origin_2018, lintott_Predicting_2022, hopkins_Galactic_2023}. 
To begin, it is reasonable to assume that ISOs should trace the same phase space density as stars, given that ISOs are produced around and ejected from individual stars, and once ejected, orbit in the same Galactic potential as the stars \citep{gaidos_Origin_2017,feng_Oumuamua_2018}.  
\citet{hopkins_Galactic_2023} and \citet{hopkins_Predicting_2024} refined this further by considering that ISOs from stars that have already died should persist in the Galaxy, and ISOs are probably more likely to be produced by higher-metallicity stars, so the local stellar population can be reweighted to include these effects in predictions for future surveys. 
Chemodynamic effects will be present in the ISO population, such as correlations between velocity and properties such as water-mass fraction and age \citep{hopkins_Predicting_2024}.
We refer to this framework as the \={O}tautahi-Oxford ISO population model.

In this work, we consider two additional, qualitatively different, effects that differentiate the dynamics of the Galactic population of ISOs from stars, expanding the scope and sophistication of the \={O}tautahi-Oxford model. 
First, when ISOs are unbound from their parent star system, they may find themselves in the birth cluster of their star. 
Such ISOs will then populate the cluster potential and, eventually, escape from it \citep{Levison_2010, Hands_2019}.
Regardless of this potential ``preprocessing'' by the birth cluster, nearly all will eventually orbit in the Galactic potential. 
Second, any material orbiting in the Galaxy with a finite spread in either initial position or velocity will generically produce tidal tails or \textit{streams}. 
This phenomenon is well-studied in the context of stellar streams produced by star clusters and dwarf galaxies disrupting in the Milky Way's potential, motivated by the discovery of stellar streams in large surveys \citep[e.g.][]{odenkirchen_Detection_2001, rockosi_MatchedFilter_2002}. 
Observed stellar streams are excellent tracers of the Galactic potential because of their low internal velocity dispersions \citep{bonaca_Stellar_2024}, and they may also retain information about their interactions with dense substructures \citep{bonaca_Spur_2019}.
The cumulative Galactic interstellar object population is often described with two broad assumptions to first order: that it is well-mixed, and that it is of uniform density.
However, ISOs will also form streams within their Galaxy. 
This was observed in simulations by \citet{portegieszwart_Oort_2021} in one realization of their ISO populations originating from the 200 nearest stars to Earth. 
Radiants of small interstellar meteors could form in this way (\citet{taylor_Discovery_1996}; \citet{baggaley_Advanced_2000}; \citet{gregg_Case_2025}), though lack confirmed observations \citep{musci_Optical_2012,froncisz_Possible_2020,hajdukova_No_2024}. 

In this work we examine the consequences of the fact that the ISO population is not a uniform background, nor even a population that traces movement akin to a subpopulation of the Galaxy's stars as assumed in most previous work, but an overlapping tapestry of streams. 
In Section \ref{sec:sims} we simulate ISO streams, noting the key features of the streams that affect their observability: their length, volume, density structure, and internal velocity dispersion. 
Then in Section \ref{sec:popmodel}, we develop a population model of the ISO streams, to predict what we expect to be the main observable consequence of ISOs' configuration in streams: how often we will see multiple ISOs from the same star or star cluster. 
As the observable volume of the Solar System is where sky surveys provide direct constraints on the local ISO number density and velocity distribution, starting in Section~\ref{sec:popmodel} we use this location in the Galaxy as our example encounter point. 

The propagation of ISO streams has many similarities to the geomorphologic properties of a major feature of Aotearoa New Zealand's southeastern landscape, the braided rivers.
Indeed, the number of ISOs in a single stream has an order-of-magnitude similarity to the number of pebbles that make up the bed of the awa Waimakariri, the largest river near \={O}tautahi/Christchurch\footnote{A length of $\sim 100\ \mathrm{km}$ and cross-sectional area $\sim (\pi/2) (100\ \mathrm{m})^2$ with pebble sizes $\sim 2\ \mathrm{cm}$ yields $\sim 10^{14}$ pebbles, perhaps consistent with an M dwarf's ISO stream.}. 
We therefore describe this population in te reo M\={a}ori in homage to the landscape that supports us and its community, he awa whiria: the braiding rivers \citep{wilkinson_Matauranga_2020,martel_He_2022,macfarlane_He_2024}.

\section{Stream Simulations}
\label{sec:sims}

A cloud of ISOs that are unbound from their progenitor system will gradually develop into a tidal stream with a component leading and a component trailing the progenitor.
The ISOs with lower angular momenta about the Galaxy will form the leading tail of the stream, and those with higher momenta form the trailing tail of the stream. 
The streams lengthen over time, and given enough time ($\gtrsim$ Gyr) will eventually wrap around the Galaxy.
A single stream that has wrapped around the Galaxy develops a braided structure if viewed in cylindrical coordinates about the Galaxy, $r$-$\phi$.
Like their namesake braided rivers, the structures form under rapid, highly fluctuating rates of discharge, from a source with a high rate of supply \citep{cant1982fluvial}.

In this section, we develop an understanding of streams' properties and how they vary, such that we can later estimate (\S~\ref{sec:popmodel}) the rate of encounters between a given stream and a star system (e.g. the Solar System) contained within the stream's volume. 
To do so, we first explore the {\em density distribution} of the stream and how it varies with the ejection velocity and age of the stream. 
All streams have complex internal density structures which vary with time --- making it critical to be able to easily pinpoint the positions and velocities of their component ISOs throughout their orbital evolution.
We represent each tidal stream as a collection of test particles orbiting in a Galactic potential, produced by a single `progenitor'. 
The progenitor may in general be a single star or a cluster of stars --- and the progenitor may by now no longer exist, either because the star has died or the cluster has dispersed. 
The test particles are initialized along the orbit of the progenitor, and to specify their orbit we need to know their 6D position and velocity at the time of ejection.
Several choices we make, particularly around how we initialize the stream, are deliberately made to keep the simulations and our interpretations of them as simple as possible.

\begin{figure}
    \centering
    \includegraphics[width=1\linewidth]{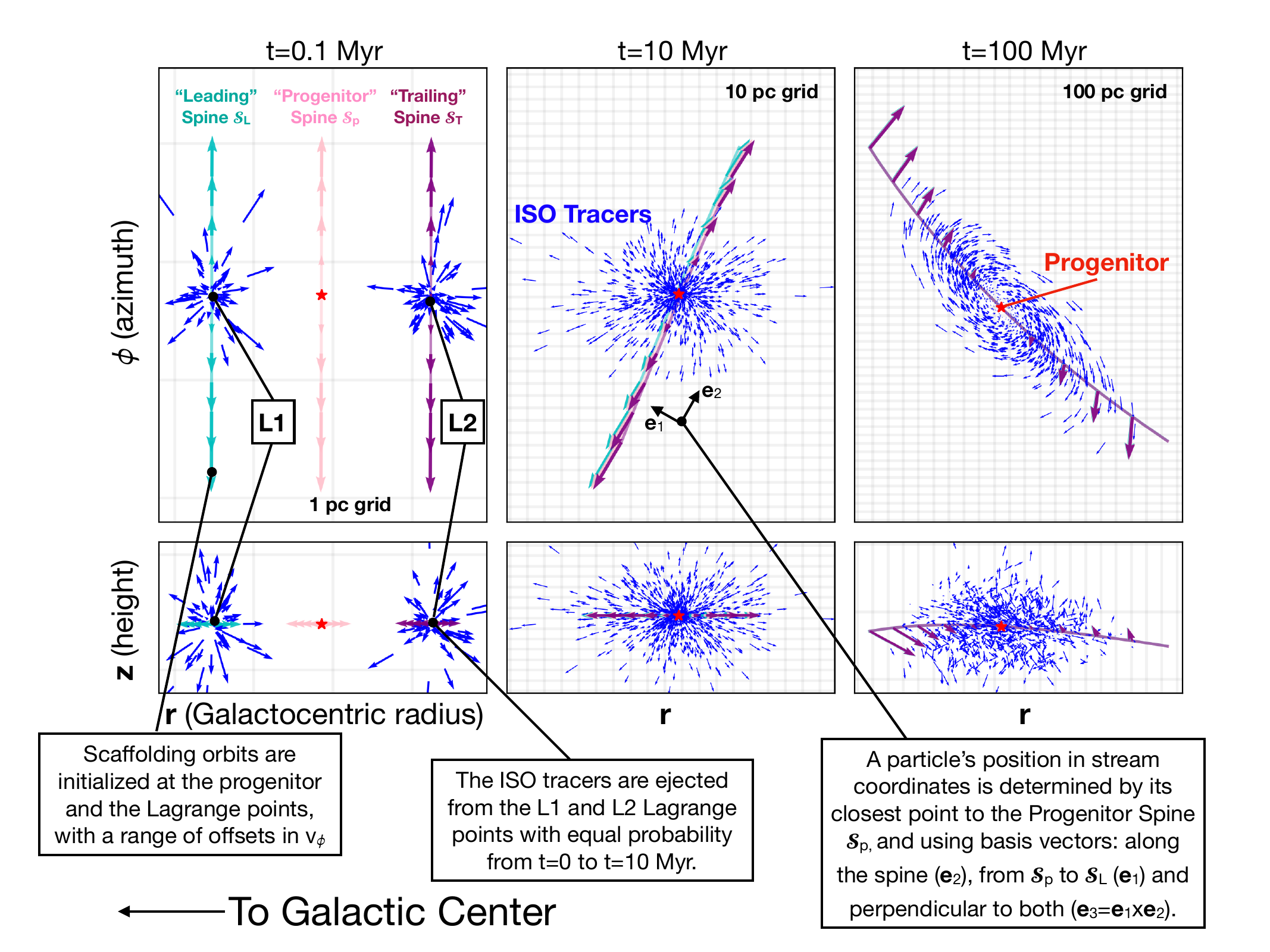}
    \caption{Initialization and early evolution of a stream. The top panels show projections in $r-\phi$ space, and the bottom panels are $r-z$ projections, where $r$, $\phi$, and $z$ are cylindrical coordinates about the Galactic center. Each column shows a different time. As time advances, the stream expands so the scale of the grid cells in each panel increases by a factor of 10. The blue arrows show the positions and velocities of ISO test particles initialized at L1 and L2. Cyan, pink, and purple arrows show a small fraction of the scaffolding orbits used to construct the stream's coordinate system. After 100 Myr, the ISOs' initial apparently-isotropic velocity distribution is replaced by a rotational pattern as each ISO particle moves about its own epicycle with a period of order $2\pi/\kappa \approx 160$ Myr. This particular stream has $M_\mathrm{prog}=1 M_\odot$, and $\sigma_0=3.1$\kms.}
    \label{fig:explainer}
\end{figure}

As is standard in the initialization of stellar streams \citep[e.g.][]{kupper_Structure_2008,kupper_More_2012,gibbons_Skinny_2014,amorisco_Feathers_2015,fardal_Generation_2015,price-whelan_Gala_2017,chen_Improved_2024}, we place newly-ejected ISOs at one of the Lagrange points of the progenitor, with respect to the Galactic potential (Figure~\ref{fig:explainer})\footnote{In the context of stellar streams, this is done because stars can only escape the progenitor if their Jacobi energy (a quantity conserved in the restricted three-body problem) is large enough, but even stars with large Jacobi energy may happen to remain physically within the cluster. 
Stars will therefore tend to leak out at the saddle points in the effective potential, namely L1 and L2 \citep{hayli_Numerical_1970}.}.
These points are offset from the progenitor by a distance
\begin{equation}
\label{eq:rL}
    r_L = \left( \frac{ G M_\mathrm{prog}}{4\Omega^2-\kappa^2} \right)^{1/3},
\end{equation}
where $G$ is Newton's gravitational constant, $M_\mathrm{prog}$ is the progenitor's mass, and $\Omega(r)$ and $\kappa(r)$ describe the local shape of the Galaxy's rotation curve at a given Galactocentric radius $r$. 
The orbital frequency is $\Omega = v_\mathrm{circ}/r$, where $v_\mathrm{circ}$ is the circular velocity at a given Galactocentric distance $r$. 
The epicyclic frequency is $\kappa = \Omega \sqrt{2(\beta+1)}$, where $\beta$ is the logarithmic derivative of the circular velocity, $d\ln v_\mathrm{circ}/d\ln r$. 
In 3D space, the two Lagrange points lie along a line from the center of the Galaxy to the progenitor, which will in general be on an orbit with non-zero vertical oscillations.\footnote{Since we primarily consider streams within the disk of the Galaxy, the relevant Lagrange points may in fact be shifted due to the tidal influence of the disk \citep[e.g.][]{heisler_Influence_1986}, but we ignore this complication for now.}

Since we are not dealing with stellar streams, but rather ISO streams, it is worth explaining our placement of ISOs at L1 and L2 in more depth. In our model, the progenitor of an ISO stream may be a single star, or a star cluster. In the latter case, the ISOs will have been ejected from their individual parent star system, but early enough and at a low enough velocity that they remain bound to the star cluster. The ISOs will then be subject to similar perturbations that allow some stars in the cluster to escape \citep{flamminidotti_Dynamical_2025}, though the process may not be perfectly analogous given the dramatically lower masses of the ISOs. Nonetheless we expect the placement of ISOs at L1 and L2 to be an excellent approximation in this particular case.
In the case of ISOs ejected from a single star, however, we need to consider two sub-cases: ISOs ejected from single encounters with a planet, or ISOs eroded from an Oort cloud (discussed more in \S~\ref{sec:ejection_mechanisms}). Erosion from an Oort cloud, in analogy to the cluster case, is likely to occur preferentially at the saddle points of the potential, i.e. we again expect L1 and L2 to be a reasonable location to place the ISOs. 
ISOs ejected from the inner parts of the star system via strong encounters with a planet, however, will be ejected essentially from the progenitor rather than the Lagrange points. 
In this case the typical velocity of ejection will be far larger than the Keplerian velocity at the Lagrange points, and so it will be unimportant that we choose that the particle was initialized at a Lagrange point instead of the progenitor, assuming that the ejections are isotropic\footnote{Depending on the details of the system, these ejections by planets may occur preferentially in the ecliptic plane of the progenitor system, which would qualitatively change the results of this section \citep{portegieszwart_Oort_2021}. However, due to the subsequent influence of heating (\S~\ref{sec:heating}), we expect that any early anisotropy in the ejection geometry will be washed out within about a Gyr.}. 
We are therefore in a situation that the only time the particle's initial location matters is when they escape at the Lagrange points, so it is reasonable to initialize all particles at the Lagrange points.

The test particles representing ISOs are placed at the two Lagrange points, at a velocity equal to the progenitor's velocity plus a random component $\delta \vec{v}$. 
Each test particle has a 50-50 chance of being placed at the L1 or L2 Lagrange point of the Galaxy-progenitor system. 
See Figure~\ref{fig:explainer} for a visual representation of this initialization, and stream coordinate system (described further below).
The three components of $\delta \vec{v}$ are drawn from a multivariate normal distribution with a diagonal covariance matrix $\sigma_0^2 I_3$. 
The velocity dispersion $\sigma_0$ is taken to be a free parameter, and $I_3$ is the 3x3 identity matrix. 

The time at which a test particle is initialized in this manner will also affect the structure of the stream \citep{amorisco_Feathers_2015}. 
We take the progenitor to produce ISOs at some rate $P(t)$, which for simplicity we assume is via an early event that releases ISOs at a high rate of supply: constant between two times $t_0$ and $t_f$, and zero at all other times.
This approximates the ISOs as produced early in the life of a planetary system via interactions with the forming planets (we consider the potential nuances of dynamical unbinding mechanisms and their variations in the rate of supply in \S~\ref{sec:ejection_mechanisms}).
In the case of star clusters on eccentric orbits about the Galaxy, this may be too simple a parametrization, since as the progenitor approaches pericenter, $r_L$ is reduced which may lead to the ejection of more material.
However, we expect the majority of streams relevant to the detection of ISOs in the Solar System to be on disk-like orbits\footnote{This follows from the fact that most stars in the Solar neighborhood are on disk-like orbits. Weighting ISO production by metallicity as in \citet{hopkins_Galactic_2023} will further reduce the importance of the halo and thick disk. 
There are however effects in the opposite direction, namely that streams on halo-like orbits suffer far less heating, so they will be denser and hence produce a disproportionately large encounter rate, if the Sun happens to lie within such a stream.}, where this effect is more modest. 
Moreover, in our approximation, $t_f-t_0$ is likely to be much shorter than any Galactic timescale, which removes any sensitivity to the exact shape of $P(t)$. 

Given a set of test particle positions and velocities at a variety of ejection times between $t_0$ and $t_f$, it is straightforward (though cumbersome) to integrate their motion forward in the Galactic potential. 
To avoid running these integrations explicitly, we instead simply ``write down'' the orbital solution following the formalism of \citet{lynden-bell_Bound_2015}, essentially a higher-order epicyclic approximation. 
We use our \texttt{LBparticles} package\footnote{\url{github.com/lbparticles/lbparticles}}, which implements an extended version of the \citet{lynden-bell_Bound_2015} formalism to include vertical motions under the influence of a disk with an arbitrary radial density profile, and corrects several minor errors in the cosine expansions relating time or azimuthal angle to corresponding internal angles (analogous to the mean and true anomaly for Keplerian orbits) (Forbes et al in prep). While this method includes vertical oscillation of the particles with a frequency that depends on galactocentric distance, it does not include the effect discussed in \citet{dehnen_Tidal_2018}, namely that the vertical oscillation frequency depends on the amplitude of the oscillation leading to tidal ribbons rather than streams. This effect is moderated provided that the vertical oscillations of particles remain well within a disk scaleheight.
We adopt the axisymmetric time-independent MW2014 potential \citep{bovy_Galpy_2015} throughout this work.

With the test particle orbits now accessible via \texttt{LBparticles}, we can evaluate the position and velocity of all test particles in the stream at any time $t$ at minimal marginal cost. 
To understand the stream's properties based on the particles' positions and velocities at a moment in time, it is useful to construct an estimator of the stream's phase space density $f(\vec{x},\vec{v})$. 
The streams generally have nontrivial six-dimensional phase space structure -- they are highly anisotropic, with over- and under-dense regions, and low sinuosity.
The stream develops over-dense regions due to the scale imprinted by the individual epicyclic motions of particles \citep{kupper_Structure_2008}. 
Moreover, dynamically cold streams will have thin substructures or ``feathering'' \citep{amorisco_Feathers_2015}. More realistic streams will have additional structures imprinted by resonances with Galaxy-scale non-axisymmetric features like the bar (\citet{pearson_Gaps_2017}), and/or compact massive structures \citep{bonaca_Spur_2019}.
In the presence of structure in the stream's phase space density, even in axisymmetric time-independent potentials, an isotropic kernel density estimator (KDE) with a kernel that does not vary from point to point may be insufficiently accurate. 
We employ a new adaptive KDE for this task, as detailed in Appendix \ref{app:kde}. 
Hyperparameters of the adaptive KDE are fixed via cross-validation, and a convergence test is shown in Appendix \ref{app:convergence}.

Another key component of our analysis is a stream coordinate system (Figure~\ref{fig:explainer}). 
We wish to know for an arbitrary position $\vec{x}$ how far ``along'' the stream we are, $\ell$, relative to the progenitor, and where we are relative to the ``core'' of the stream in a cross-sectional slice through the stream at a location $\ell$ along the stream. 
To construct such a coordinate system, we initialize three additional sets of particle orbits at time $t_0$: one at each of the Lagrange points, and one at the progenitor's position. 
Each orbit begins with the progenitor's velocity plus an adjustment to the particle's $v_\phi$, that is, the particle's azimuthal velocity in the cylindrical coordinate system aligned with the disk potential.
These ``scaffolding'' orbits trace out the path of the stream, by covering the spread in angular momenta of the test particles representing ISOs. 
Each set of scaffolding orbits is regularly spaced in $\delta v_\phi$ from $-5\sigma_0$ to $5\sigma_0$. 
The coordinate $\ell$ along the stream is then defined as
\begin{equation}
    \ell(\delta v_\phi) = \int_0^{\delta v_\phi}|(1/\epsilon)(\vec{\mathcal{S}_p}(\delta v_\phi' + \epsilon) -  \vec{\mathcal{S}_p}(\delta v_\phi' ))| d\delta v_\phi',
\end{equation}
where $\vec{\mathcal{S}_p}$ is a set of 3 univariate splines interpolated from the $x$, $y$, and $z$ Cartesian coordinates of the regularly-spaced (in $\delta v_\phi$) scaffolding orbits. 
In other words, for the purposes of this equation we are using $\delta v_\phi$ as an arbitrary coordinate and stepping along the path of the scaffolding orbits in Cartesian coordinates to build up the distance $\ell$.
Here $\epsilon$ is chosen to be small compared to $\sigma_0$ -- in practice we choose $10^{-4}$ \kms.
We use $\vec{\mathcal{S}}_p$ to refer to the splines constructed from the scaffolding orbits initialized at the progenitor, and $\vec{\mathcal{S}}_L$ and $\vec{\mathcal{S}}_T$ to refer to the splines from the scaffolding orbits initialized at the inner and outer Lagrange points respectively (corresponding to the leading and trailing streams\footnote{The velocity dispersion of ISOs is often large enough relative to the Keplerian speed at $r_L$ for the single-star case that the leading part of the stream and the trailing part of the stream both actually have contributions from both L1 and L2. Nonetheless we stick with the ``Leading'' and ``Trailing'' nomenclature for the spines.}). 
For a similar approach, see \citet{bovy_Dynamical_2014}.

The two coordinates perpendicular to $\ell$ are constructed using $\vec{\mathcal{S}}_L$ and $\vec{\mathcal{S}}_T$. 
At a given location along the stream, the nearest points to the leading and trailing streams are found numerically (in general the $\delta v_\phi$ coordinate of the leading or trailing stream at this point will not be the same as the $\delta v_\phi$ coordinate along $\vec{\mathcal{S}}_p$). 
Typically these distances will be of order $r_L$. 
Because of the finite range of $\delta v_\phi$'s used in constructing the scaffolding orbits, and the fact that sometimes (mostly just for cluster progenitors) the Galactic shear $\Omega r_L$ is larger than $\sigma_0$, the leading and trailing streams will be well-separated. 
We therefore need our procedure to work if only one of the sets of scaffolding orbits is present at a given location. 
A vector is then drawn from either the central scaffold to the trailing stream, or the leading stream to the central scaffold at this position of closest approach. 
This direction is used as $\hat{e}_1$ in the stream-centered coordinate system. 
Via the choice of sign above, this vector will be similar to $\hat{r}$. 
We choose $\hat{e}_2$ to lie along the stream, i.e. the local location of $\hat{\ell}$, such that positive values of this second coordinate correspond to larger values of $\delta v_\phi$. 
Finally the third coordinate, similar to $\hat{z}$, is set according to the right-hand rule $\hat{e}_3 = \hat{e}_1 \times \hat{e}_2$.

\begin{figure}
    \centering
    \includegraphics[width=\linewidth]{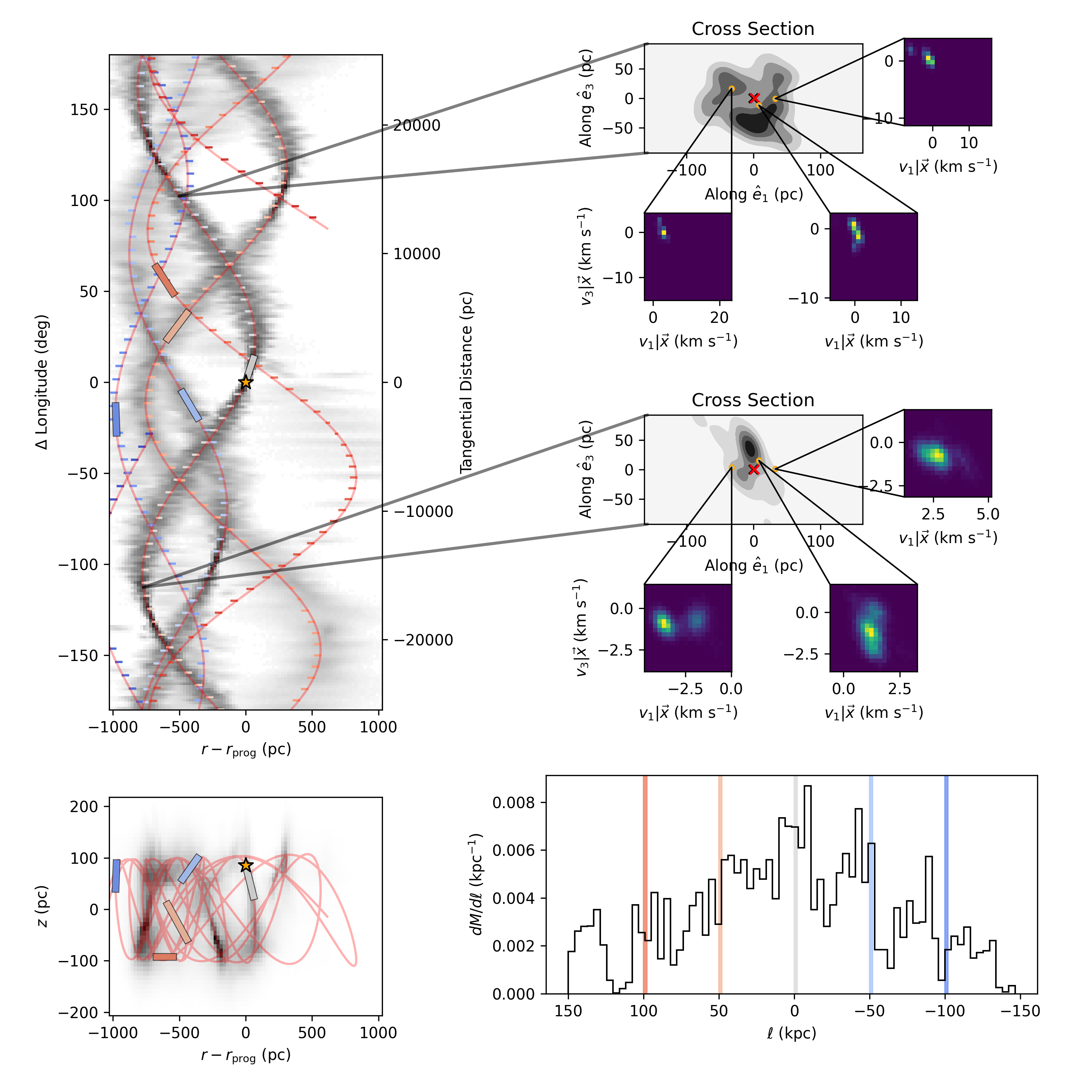}
    \caption{The structure of an interstellar object stream 10 Gyr after its formation. 
    For a progenitor star on a Sun-like orbit, $10^4$ particles are generated at the star's Lagrange points with a velocity dispersion of $3.1\ \mathrm{km}\ \mathrm{s}^{-1}$, and evolved for 10 Gyr in a Milky Way-like potential, where they wrap the Galaxy multiple times. 
    The top left panel shows the estimated density of the stream in $r-\phi$ coordinates centered on the progenitor (green star). 
    The sequence of ``scaffolding" orbits $\vec{\mathcal{S}}_P$ is shown as a red line, with ticks every kpc indicating the direction of $\hat{e}_1$. 
    The lower left panel shows the corresponding $r-z$ view of the stream's density (grey) and the scaffolding orbits (red). 
    On the right, two example locations along the stream connected by black lines the $r-\phi$ diagram display the density of a cross section of the stream (grey), centered on $\vec{\mathcal{S}}_P$ (red cross). 
    For four points in each of these cross sections, the velocity structure of the stream is shown in adjacent plots (linear particle histogram in colourmap). 
    Finally, the lower-right panel shows the density of the stream per unit length along the stream $dM/d\ell$, with thin colored regions showing the $\ell$ range covered by the 5 rectangles in the $r$-$\phi$ and $r$-$z$ diagrams.} 
    \label{fig:rphi}
\end{figure}

A snapshot of the density structure of one particular stream that has multiple wraps around the Galaxy is shown in Figure \ref{fig:rphi}.
The positions and velocities of a set of $10^4$ test particles evolved for 10 Gyr are used to construct a 6D cross-validated adaptive kernel density estimate for the phase space density of stream particles $f$. 
The particles are initialized with a velocity dispersion $\sigma_0=3.1\ \mathrm{km}\ \mathrm{s}^{-1}$ and with a progenitor mass $M_\mathrm{prog}=1\ M_\odot$. 
The stream's structure in $r-\phi$ coordinates centered on the progenitor's position resembles the growth and mobility of a river in its braidplain, he awa whiria, as the stream spreads and wraps around the galaxy, with each ISO tracer following a slightly different precessing orbit\footnote{Orbits around a galaxy are generally not closed because the period of radial oscillation ($2\pi/\kappa$) is not the same as the period of the orbit around the galaxy ($2\pi/\Omega$), so the particle traces out a Rosette pattern.} around the Galaxy. 
At the same time, each ISO is oscillating vertically at a somewhat higher frequency due to the self-gravity of the disk. 
The main structure of the stream is traced well by the scaffolding orbits $\vec{\mathcal{S}}_P$, which are shown as red curves in the left panels of the Figure. 
The gray histograms in these panels are linearly scaled and populated by drawing $10^7$ particles from the KDE and placing them in one of 240 (in the $\phi-$direction) by 90 (in the $r-$direction) bins, and one of 90 by 80 (in the $z-$direction) bins respectively. 
In other words, these plots are projections of the density, $\int \rho dz$ and $\int\rho d\phi$, respectively.

At several points along the stream, we show the density of the stream in cross-section. 
Starting from the scaffolding orbits, we evaluate the density on a grid aligned with $\hat{e}_1$ and $\hat{e}_3$. 
Since these plots are slices rather than projections, no integration is necessary and evaluation on a grid is efficient. 
Interestingly, the shape of the cross-section is not necessarily aligned with $\hat{e}_1$ and $\hat{e}_3$, since $\vec{\mathcal{S}}_T$ and $\vec{\mathcal{S}}_L$ are not strictly interior or exterior to $\vec{\mathcal{S}}_p$. 
This is visible as the direction of $\hat{e}_1$, shown by the tick marks along the scaffolding orbit in the top left panel of Figure \ref{fig:rphi}, changing directions. 
Within the cross sections, we select a few points at random and show projections of their velocity distributions $\int f(\vec{v} | \vec{x}) dv_2$ conditional on their location and projected through $\hat{e}_2$. 

Finally, we show the line density of the stream, i.e. the mass per unit length along the scaffolding orbits. 
For this calculation, $10^5$ points are drawn from the marginal distribution $p(\vec{x}) = \int f d\vec{v}$. 
Each point is then assigned to a location $\ell$ along the stream by finding the $\ell$ that minimizes the distance between the point and $\vec{\mathcal{S}}_p$. 
These values of $\ell$ are then plotted as a histogram. 
\begin{figure}
    \centering
    \includegraphics[width=0.95\linewidth]{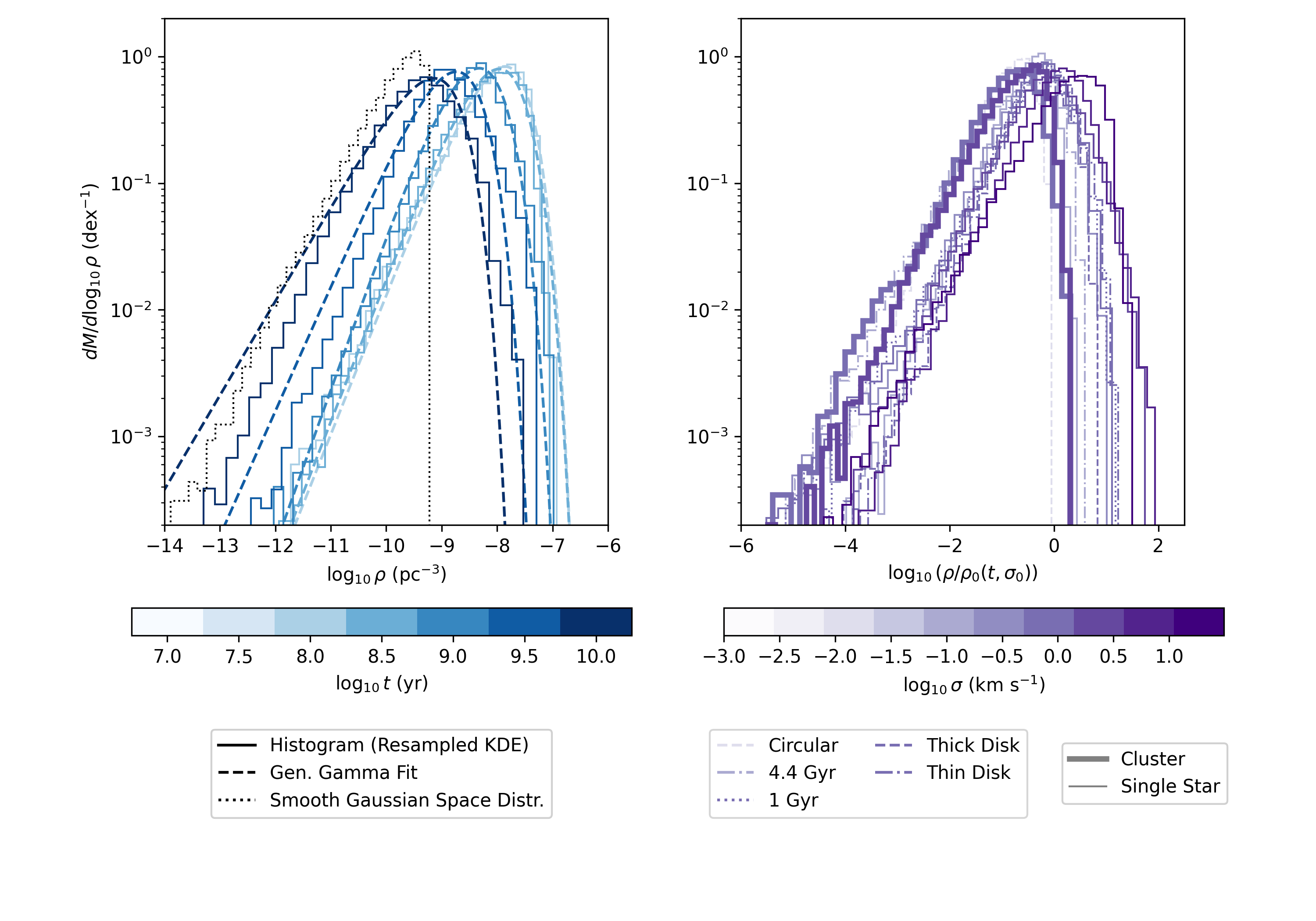}
    \caption{Mass-weighted density distribution of ISO streams. The left panel shows the evolution of one stream (with $\sigma_0=3.1\ \mathrm{km}\ \mathrm{s}^{-1}$ and $M_\mathrm{prog} = 1\ M_\odot$), and the right panel shows the PDFs for a wide range of streams, where the density is normalized to $\rho_0$. In the left panel we also show the generalized gamma fits to each time's PDF as dashed lines. For the $t=10\ \mathrm{Gyr}$ line we also show, as a black dashed line, the distribution of densities if the ISOs were distributed in a 3D gaussian in space.}
    \label{fig:densityPDF}
\end{figure}

\begin{figure}
    \centering
    \includegraphics[width=0.95\linewidth]{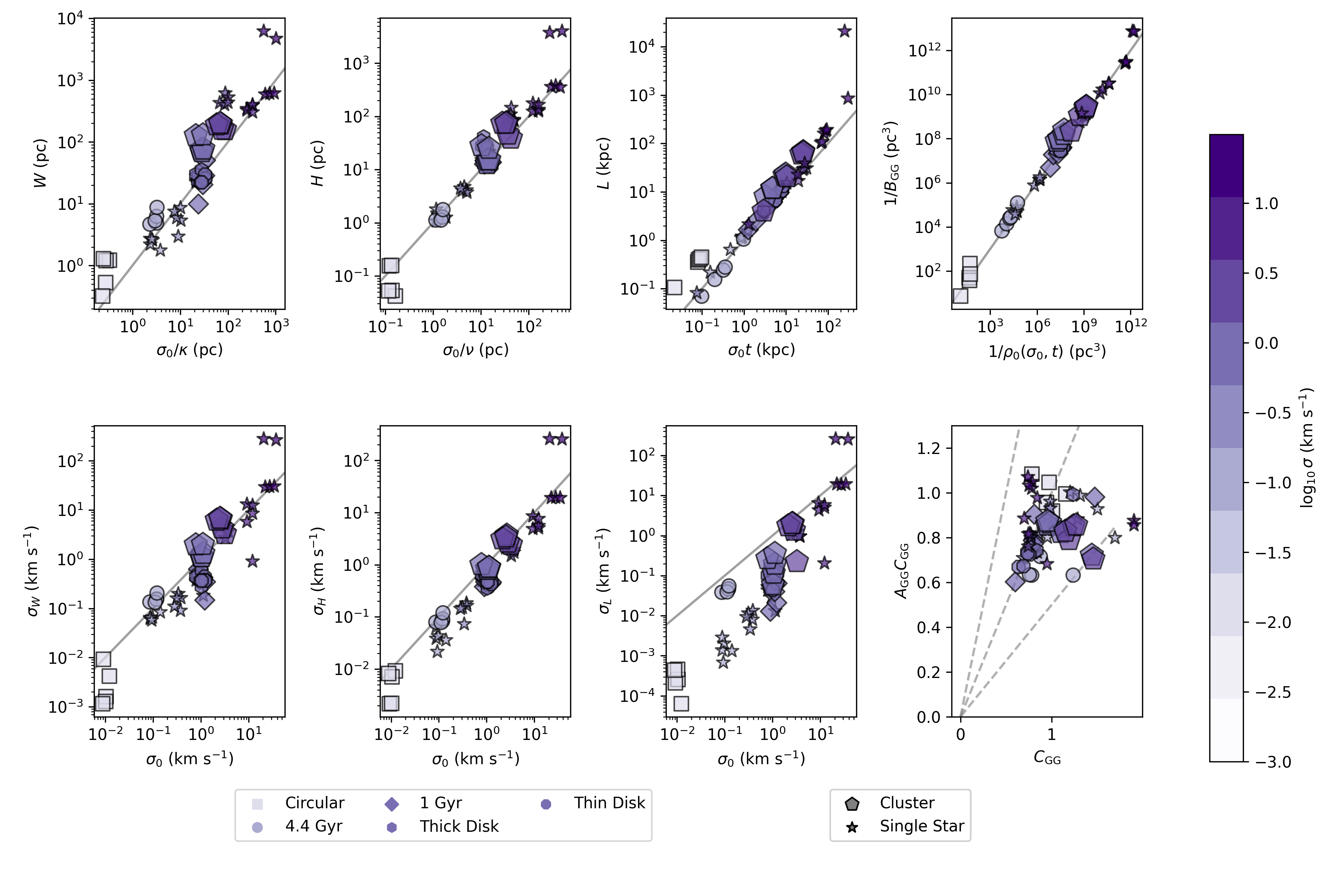}
    \caption{Summary of ISO stream morphologic properties. The top four panels compare the width, height, length, and peak density of a set of simulated streams to rules of thumb for each of these quantities. In the bottom row, the velocity dispersion along the 3 different axes of the stream is compared to the initial velocity dispersion $\sigma_0$. The bottom right panel shows the shape parameters of the density distribution. In each panel, 5 points are shown for each stream, corresponding to 5 randomly-selected times between 0 and 10 Gyr. For the most part we expect streams to follow the 1:1 line shown in these panels, and for the most part they do, over 3+ orders of magnitude. The exception is $\sigma_L$, the velocity dispersion along the stream direction, which is typically far less than $\sigma_0$, since a particle's position in the stream is primarily determined by its velocity in this direction. Note that for plotting clarity we have included a small random scatter of 0.07 dex on the x-coordinates of points where the x-axis is $\sigma_0$ times a constant, since many points have identical values. In the last panel comparing the shape parameters of the generalized gamma distribution, we include dashed grey lines at $y=x/2$, $y=x$, and $y=2x$.}
    \label{fig:streamprops}
\end{figure}

Remarkably, the density of points tends to be greatest fairly near the progenitor of the stream --- despite the fact that in this example model the progenitor has not produced any new ISOs in 10 Gyr, since its initial high-discharge/high rate of supply event. 
This density effect does not appear to be particular to this particular time or stream.
It seems to arise from the fact that the modal angular momentum of particles ejected by the stream is the same as the angular momentum of the progenitor itself. 
This overdensity of stream material around the progenitor is superficially similar to the proposed halo of $\sim 10^7$ ISOs gravitationally bound to the Sun \citep{penarrubia_Halo_2023}, with this number set by an equilibrium between new gravitational captures and tidal removals.
However, that ISO halo calculation did not include the effects of streams, and our calculations do not include the self-gravity of the progenitor. 
Depending on $\sigma_0$ and the age of the stream, the stream overdensity may be much larger or smaller than the halo of ISOs.
There are also around a dozen other peaks in the line density distribution, which likely correspond to pericenter and apocenters of the scaffolding orbits, and extrema of the vertical motion of the stream particles.  

We simulate a variety of streams to understand how their properties (e.g. length, width, density distribution) scale with input properties (e.g. $\sigma_0$, progenitor orbit, and progenitor mass). 
For most of our sample of simulated streams, the progenitor of each stream is a star on a Sun-like orbit, with each stream simulating a different $\sigma_0$. 
Two streams with cluster progenitors are simulated. 
Additionally, we explore the effect of orbital eccentricity by simulating two streams with progenitor orbits closer to circular: the thin-disk-like stream (labeled ``Thin Disk'') has a velocity difference from circular $0.31$ that of the Sun's deviation from circular, and the stream labeled ``Circular'' is just on a circular orbit. A stream with triple the Sun's peculiar velocity is also simulated and labelled ``Thick Disk''.
Finally, the streams labeled ``1 Gyr'' and ``4.4 Gyr'' extend the time for high rate of supply of the ISOs, setting $t_f-t_0=1\ \mathrm{Gyr}$ or $4.4$ Gyr respectively, whereas all other streams have the short supply time of $t_f-t_0=10\ \mathrm{Myr}$.

We are particularly interested in the mass-weighted density PDF of the stream (Figure~\ref{fig:densityPDF}), since the distribution will be crucial in the next section. 
We denote this quantity $p(\log_{10}\rho) \equiv d M/d\log_{10}\rho$: the distribution of mass in the stream per unit log density. 
Equivalently, this is the distribution of log-densities within which a randomly drawn ISO from the stream will be present. 
We normalize this distribution so that $\int p(\log_{10}\rho) d\log_{10}\rho = 1$. 
Examples of the density PDF are shown in Figure~\ref{fig:densityPDF}. 
In the left panel, a single stream's PDF is shown at a logarithmically spaced sequence of times from $10^8$ years to $10^{10}$ years. 
On the right, the density PDF is shown for a wide variety of streams (detailed below), with the density normalized to $\rho_0(t,\sigma)$, to be defined momentarily.

Figure~\ref{fig:densityPDF} demonstrates that the density PDF is well-fit by a generalized gamma distribution, namely a powerlaw with an exponential cutoff in $\rho^{C_\mathrm{GG}}$ (as opposed to just $\rho$ for an ordinary gamma distribution). 
The full PDF is
\begin{equation}
    p(x) = \frac{|C_\mathrm{GG}|}{\Gamma(A_\mathrm{GG})} (x/B_\mathrm{GG})^{A_\mathrm{GG}C_\mathrm{GG} -1 } \exp\left( -(x/B_\mathrm{GG})^{C_\mathrm{GG}} \right),
\end{equation}
where $A_\mathrm{GG}$ and $C_\mathrm{GG}$ are shape parameters, $B_\mathrm{GG}$ is the overall scale corresponding to the location of the exponential cutoff, and the first factor just normalizes the distribution. 
$\Gamma(x) = \int_0^\infty t^{x-1} e^{-t} dt$ is the Gamma function. 
Based on Figure \ref{fig:densityPDF}, most of the time the mass-weighted distribution of $\rho$ can be fit with $A_\mathrm{GG} \sim C_\mathrm{GG} \sim 1$, and $B_\mathrm{GG} = \rho_0(t,\sigma_0) \equiv \left[ (2.5 \sigma_0)^3 \kappa^{-1} \nu^{-1} t \right]^{-1}$. 
This scaling is plausible, based on the rule of thumb that the width of a stream is $\sim \sigma_0/\kappa$ \citep{carlberg_Subhalo_2023}, its height is $\sim \sigma_0/\nu$, and its length is $\sim \sigma_0 t$. 
The peak of the density distribution will then occur at the mass of the stream divided by the volume, since we have normalized the mass to $1$, $\rho \sim 1/V$, and the volume $V \sim \sigma_0^3 \kappa^{-1}\nu^{-1} t$.

To test whether our streams obey these relationships, we measure their width, height, and length (Figure \ref{fig:streamprops}; a given stream has the same color in both Figure \ref{fig:streamprops} and the right panel of Figure \ref{fig:densityPDF}).
We also measure their velocity dispersions, and the best-fit generalized gamma distribution parameters, at a variety of times randomly selected between 0 and 10 Gyr. 
The width, height, and length are estimated by drawing $10^4$ samples from the KDE. 
Each sample's $\ell$ coordinate is found by minimizing the distance to the scaffolding orbit $\vec{\mathcal{S}}_p (\ell)$. 
The sample's position is then projected onto the stream coordinate system $e_1(\ell)$, $e_2(\ell)$, and $e_3(\ell)$, yielding a position relative to the stream. 
The mean velocity at that location is estimated by drawing an additional 10 samples from the conditional distribution $p(\vec{v}|\vec{x})$. 
The offset of the particle's velocity from this mean is then projected into the stream coordinate system. 
We then define the width $W$ as the standard deviation of the samples' $e_1$ coordinates, the stream's height $H$ as the standard deviation of the samples' $e_3$ coordinates, and the stream's length $L$ as the standard deviation of the samples' $\ell$ coordinates. 
These lengths are compared to $\sigma_0/\kappa$, $\sigma_0/\nu$, and $\sigma_0 t$ respectively. 
We find reasonable agreement over four orders of magnitude, which explains why the best fit $1/B_\mathrm{GG}$ values correspond reasonably well to $\rho_0$.

The velocity dispersions along the width and height of the stream, which we call $\sigma_W$ and $\sigma_H$, correspond well to $\sigma_0$. 
This justifies our use of $\sigma_0$ in the comparison to the stream's physical dimensions. 
The velocity dispersion along the direction of the stream, $\sigma_L$, is considerably lower than $\sigma_0$ for most values of $\sigma_0$. 
This is because the set of particles at a particular location along the stream are there specifically because of their velocity along the path of the stream. 
Eventually, $\sigma_L$ does approach $\sigma_0$, when the stream begins to wrap around the Galaxy and intersect itself.

\section{Population Model}
\label{sec:popmodel}

\begin{figure}
    \centering
    \includegraphics[width=0.9\linewidth]{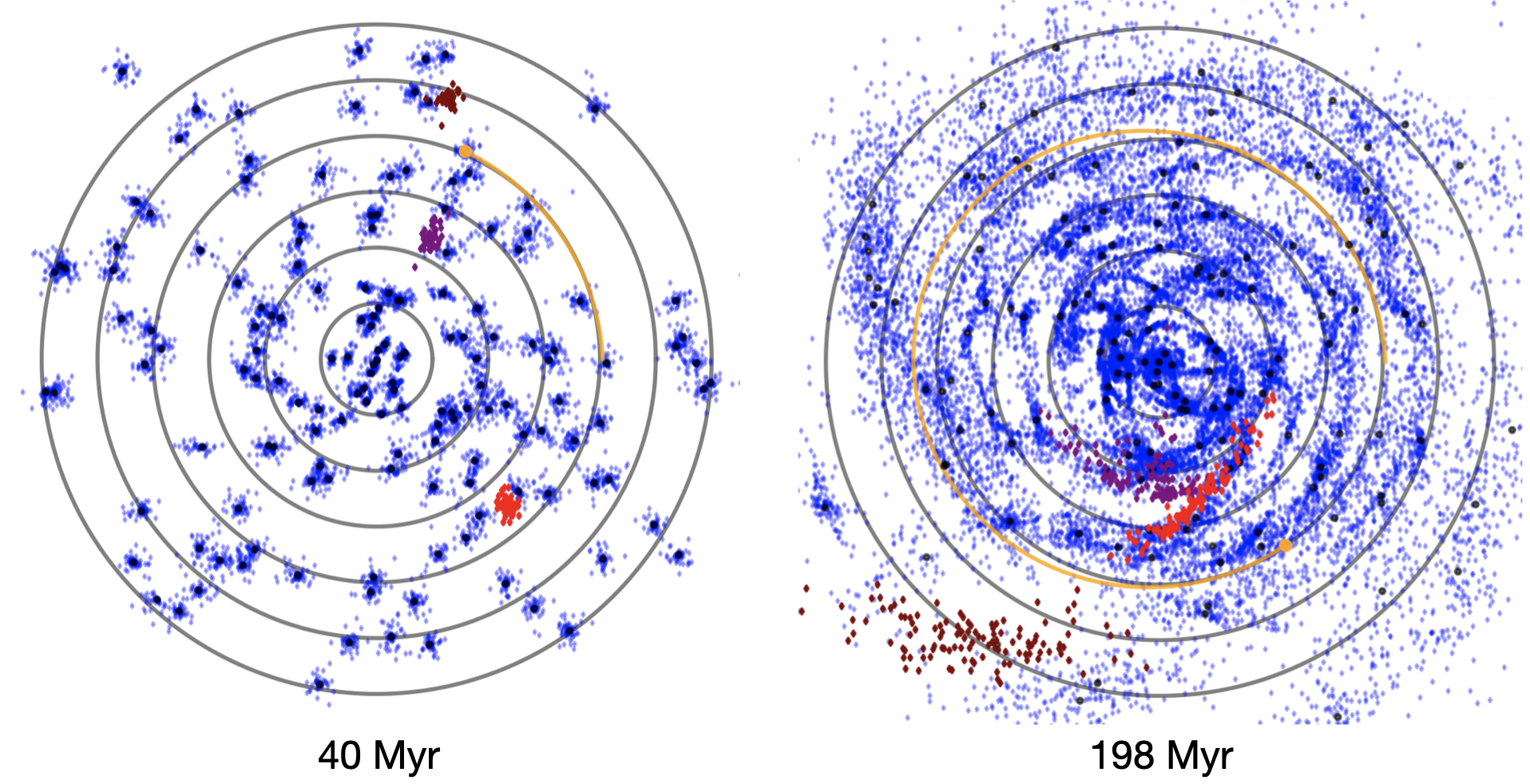}
    \caption{Interstellar object streams across the Galaxy. To visualize the wide range of streams that may intersect the Sun at any given moment, we show a small representative population of 160 progenitors (the black points) with 120 ISO tracer particles per progenitor shown in blue. This is a top-down view of the Galaxy, and the circles show circles of constant galactocentric radius in intervals of 2 kpc. The progenitors are drawn from a uniform distribution in radius. Three streams are highlighted for demonstration in other colors (purple, red, and maroon). The Sun's orbit is shown in orange, and the streams here have evolved for 40 Myr (left) and 198 Myr (right). The orbits of the particles are highly idealized, since the potential (MW2014 from \citet{bovy_Galpy_2015}) is axisymmetric and does not evolve with time. The most dramatic effect missing from this visualization is likely that particles within the region affected by the bar ($r \lesssim 4$ kpc; \citet{bland-hawthorn_Galaxy_2016}) will be redistributed by resonances with the bar's orbit (e.g., \citet{athanassoula_Bars_2013}).  }
    \label{fig:galaxy_year_frame}
\end{figure}

The population of ISOs observed in the Solar System will originate from a variety of streams, each of which will have its own velocity dispersion, lifetime, heating history, and progenitor (cf. Figure~\ref{fig:galaxy_year_frame}). 
Given our understanding of the fundamental properties of streams, we now aim to predict how streams will affect the population of ISOs encountering our Solar System. 
At the most basic level, we would like to know how many streams contribute to the populations of ISOs visiting the Solar System.
If this number is very large, each ISO is likely to come from a different stream, whereas if a few streams dominate, we would see ISO ``siblings,'' that is, multiple ISOs from the same stream. 
In the most extreme case where multiple ISOs come from a dynamically-cold stream, they may appear as ``twins'': ISOs with very similar incoming velocities and radiants\footnote{As far as our models are concerned, any binary ISO would be considered here as a single ISO for its dynamics in the Galaxy and encounter rate with the Sun; its components would not count as siblings in the sense we use throughout this paper.}. 
In order to compare the relative properties of each population model in this section, we will normalize the total rate of ISOs entering a sphere of $q_\mathrm{max} = 5$~au around the Sun to $10\ \mathrm{yr}^{-1}$. 
This is a broad-brush approximation to the estimates of the background number density of ISOs and the ability to find ISOs of a decade-long survey such as LSST. 
A more detailed accounting would depend on ISO orbits, size distributions, and observing strategy (\citet{dorsey_Visibility_2025}).

At an order-of-magnitude level, the number of streams contributing to the ISOs encountering the Solar System is just the ``volume'' of a typical stream\footnote{Streams do not have sharp edges, but there is a well-defined peak in the density distribution (Figure \ref{fig:densityPDF}), so volume is a fuzzy but meaningful quantity.} multiplied by the number density of progenitors, $n_\mathrm{prog}$. 
Figure \ref{fig:streamprops} demonstrates that streams have a width of order $\sigma/\kappa$, a height of order $\sigma/\nu$, and a length of order $\sigma \tau$, where $\tau$ is the stream's age. 
The number of streams is therefore of order
\begin{equation}
\label{eq:Nstreams}
    N_\mathrm{streams} \sim \pi n_\mathrm{prog} \sigma^3 \tau \kappa^{-1}\nu^{-1} \sim 2\times 10^5 \left(\frac{\sigma}{1\ \mathrm{km}\ \mathrm{s}^{-1}}\right)^3 \left(\frac{n_\mathrm{prog}}{0.1\ \mathrm{pc}^{-3}} \right) \left( \frac{\tau}{5\ \mathrm{Gyr}}\right),
\end{equation}
where we have adopted Solar neighborhood values for $\kappa \approx \sqrt{2}\ 220\ \mathrm{km}\ \mathrm{s}^{-1} / (8\ \mathrm{kpc})$ and $\nu \approx \sqrt{4\pi G (0.1\ \mathrm{M}_\odot\ \mathrm{pc}^{-3})}$. 

If we expect to observe $N_\mathrm{ISOs} \sim 100 \pm 1\ \mathrm{dex}$ ISOs over the course of the next decade, the large prefactor, $2\times 10^5$, in our estimate of $N_\mathrm{streams}$ tells us that for the reasonable values we have adopted here, $N_\mathrm{ISOs} \ll N_\mathrm{streams}$, in which case we might expect streams to have relatively little effect on the observed distribution of ISOs. 
Several distributional effects may act to soften this conclusion, however. 
First, $n_\mathrm{prog}$ may be much lower than the fiducial value we have adopted if ISOs essentially originate in clusters, rather than from individual stars. 
This scenario is reasonably likely, given both that most stars form in clusters, and that we expect most ISOs to be ejected early in the lifetime of a star at velocities low enough to be bound by the cluster. 
In this case, the effective number density of progenitors may be reduced by a factor of order $\langle M_* \rangle/\langle M_\mathrm{cluster} \rangle$, the ratio of the average mass of individual stars drawn from the IMF to the average mass of the initial cluster mass function. 
{\bf To distinguish between ISOs from the same star, and ISOs from the same cluster, we refer to the former as siblings, and the latter as cousins.} 
The number of ISO siblings or cousins varies substantially depending on the parameters of the population model, and therefore may be used to distinguish between them, if two ISOs can be determined to be siblings or cousins observationally.

The estimate of Equation \ref{eq:Nstreams} also assumes that all streams are sufficiently similar that we can use a single value of each quantity, whereas the population of streams should have a wide range of ages and may be characterized by a wide range of velocity dispersions. 
Even within a single stream, Equation \ref{eq:Nstreams} assumes that the density of ISOs is uniform, whereas, as we saw in the previous section, ISO streams likely have a wide range of densities encompassed in their ``volumes'' (Figure~\ref{fig:rphi}, see also \citet{portegieszwart_Origin_2018}). 
Any high-density, low-velocity dispersion, young contribution to the stream (\S~\ref{sec:ejection_mechanisms}) may therefore have an outsized effect, producing ISO siblings or twins. 
In the other direction, streams' velocity dispersions will increase over time as they encounter perturbers along their orbits in the Galaxy.

To account for these effects, we construct a Monte Carlo model to draw ISOs from the Galactic population. 
We assume that each progenitor generates a number of ISOs proportional to the mass of the progenitor, and that these ISOs form a stream. 
For now, we neglect the complication that each stream may be composed of multiple components. 
Rather, we take each stream to be isothermal, characterized by a single age\footnote{Note that this is slightly different than our setup in the simulations of individual streams, where we initialized particles at a range of times --- for the population models, we assume the ISOs are released in a single burst for simplicity.} $\tau$ that we take to be equal to the time between the birth of the star or cluster and the present day, a velocity dispersion $\sigma$, a progenitor velocity $\vec{v}_\mathrm{prog}$, a progenitor mass $M_\mathrm{prog}$, and three numbers characterizing the stream's generalized gamma density distribution (Table~\ref{tab:popmodels}).
Draws from the Monte Carlo distribution proceed as follows. 
First, a set of progenitor properties is drawn. 
The mass of the progenitor is drawn from a \citet{kroupa_Variation_2001} initial mass function (IMF) for the single star models, or for the cluster cases, a power law distribution between $M_\mathrm{min}$ and $M_\mathrm{max}$ with slope $\beta$ {so that the number of clusters per unit mass $dN/dM \propto M^\beta$ for $M_\mathrm{min}<M<M_\mathrm{max}$. 
The typical velocity of the stream in the local standard of rest is taken to be a 3D isotropic Gaussian with 1D velocity dispersion $\sigma_\mathrm{prog}$; this distribution can be made more realistic in the future following \citet{hopkins_Predicting_2024}. 
It is important to note that the distribution of velocities in the Solar neighborhood is likely to be affected by resonances with the bar (e.g. \citet{moreno_Effect_2021}; \citet{dillamore_Radial_2024}). Our adopted density structure for each stream (see \S \ref{sec:sims}), which may also be affected by the bar (e.g. \citet{pearson_Gaps_2017}), may therefore need to be altered for certain parts of the local velocity distribution. We expect that this will be a second-order effect (that is, adopting a velocity distribution that more closely matches the Gaia data following \citet{hopkins_Predicting_2024} will be more important), but given that the origins of the structures in velocity space are not well-understood, these effects may matter more than we expect.
The age of the stream is drawn uniformly from 0-12 Gyr, reflecting the roughly constant star formation history expected for the Milky Way \citep[e.g.][]{schonrich_Chemical_2009}. 
Note that the progenitor star or cluster does not need to still be extant for the ISO stream to persist, so it is the star formation history, not the age distribution of current stars, that matters \citep{lintott_Predicting_2022,hopkins_Galactic_2023}.

Each stream is then assigned a value of $\sigma_\mathrm{birth}$, the spread of ISO velocities at the moment of their ejection. 
These values are drawn from a lognormal distribution characterized by a median $\sigma_\mathrm{birth,med}$ and width $s_\mathrm{birth,stdev}$ given in dex, with a floor at the Keplerian velocity of the progenitor star at its tidal radius with respect to the Galaxy,
\begin{equation}
v_T = \sqrt{\frac{G M_\mathrm{prog}}{r_L}} = 0.055\ \mathrm{km}\ \mathrm{s}^{-1}\ \left(\frac{M_\mathrm{prog}}{1 M_\odot}\right)^{1/3}.
\end{equation}
Here we have re-used $r_L$ from equation \ref{eq:rL}, and adopted Solar Circle values for $\Omega = (220\ \mathrm{km}\ \mathrm{s}^{-1})/(8\ \mathrm{kpc})$ and $\kappa=\sqrt{2}\Omega$.

We then assume that the stream is subject to dynamical heating as it evolves in the Galactic disk. 
As a first approximation, we assume that the heating follows a power law index $\alpha_H$ so that after some time $t$ a stream which began with a velocity dispersion $\sigma_\mathrm{birth}$ ends up with a velocity dispersion
\begin{equation}
    \sigma = \left(\sigma_\mathrm{birth}^{1/\alpha_H} + \mathcal{H}\frac{t}{14\ \mathrm{Gyr}}  (35\ \mathrm{km}\ \mathrm{s}^{-1})^{1/\alpha_H} \right)^{\alpha_H}.
\end{equation}
A constant heating rate, i.e. $d\sigma^2/dt = \mathrm{const.}$ would correspond to $\alpha_H=0.5$. 
If the observed age-velocity dispersion relation of stars in the Solar neighborhood is purely the result of heating by the disk, then $\alpha_H$ could be read off from the slope of this relation, and $\mathcal{H}\approx 1$ so that over 14 Gyr, the velocity dispersion would increase to $\approx$ 35 km s$^{-1}$. 
We note however that there are alternative explanations for the age-velocity dispersion correlation, including the possibility that older stars were born dynamically hotter \citep[e.g.][]{bournaud_thick_2009, forbes_evolving_2012}. 
In this case, $\mathcal{H} \la 1$. 

We assume then that the density distribution associated with each stream is given by a generalized gamma distribution, as discussed in \S~\ref{sec:sims} (cf. Fig.~\ref{fig:densityPDF}). 
The shape parameters are drawn from normal distributions, as is the scale parameter $B_\mathrm{GG}/\rho_0$. 
Each of these normal distributions is truncated at 0.01 to ensure the distribution parameters remain positive.

For each sampled stream we now draw a rate $\mathcal{R}_i$ of encounters with the Solar System, where as we chose earlier, the ISO has some pericenter less than $q_\mathrm{max}= 5\ \mathrm{au}$. 
The rate of encounters with the stream is 
\begin{equation}
    \mathcal{R}_i = \left[\pi q_\mathrm{max}^2 \left(1 + \frac{2 G \cdot 1 M_\odot}{q_\mathrm{max} v_\mathrm{rel}^2}\right)\right] |\vec{v_\mathrm{rel}}|\ \rho_i\ \mathcal{N}\ \frac{M_\mathrm{prog}}{1\ M_\odot}
\end{equation}
The first factor, in square brackets, is the velocity-dependent cross section for an ISO to interact with the Solar System including gravitational focusing from the Sun. 
The rate is this cross section times the relative speed of the ISOs and the Sun, multiplied by the density of ISOs. 
The velocity is estimated as
\begin{equation}
    \vec{v}_\mathrm{rel} = \left((\vec{v}_\mathrm{stream} - \vec{v}_\odot + \sigma \mathcal{Z}_{2,i})  \right),
\end{equation}
where $\vec{v}_\odot = (-11.1, 12.4, 7.25)$ km s$^{-1}$ is the Sun's velocity in the LSR \citep[adopted from][]{bland-hawthorn_Galaxy_2016}, and the third term accounts for the additional random velocity from the stream's internal velocity dispersion. 
In particular $\mathcal{Z}_{2,i}$ is a draw from a 2D gaussian with a covariance matrix equal to the identity matrix. 
This 2D vector is then oriented randomly in 3 dimensions, to model the fact that each stream's velocity dispersion is large and comparable to $\sigma_0$ perpendicular to the stream's scaffolding orbits, but small along the direction defined by those orbits. 
The density $\rho_i$ is drawn from the mass-weighted density function of the stream as modelled by the generalized gamma distribution, accounting for the fact that the Sun could be at any of a range of densities within the stream. 
Since we have normalized the stream density distribution such that it integrates to a single ISO, we need to multiply by the typical number of ISOs per Solar mass, which we denote $\mathcal{N}$. 
For now we set this equal to $10^{16}$, but to facilitate model comparison, we will adjust it for each population model such that the total rate of ISO encounters with the Solar System is $10\ \mathrm{yr}^{-1}$ (see below).

The Sun is not more likely to encounter a dense part of an ISO stream than a less-dense part, so to draw a fair sample of rates we would need to sample $\rho_i$ from the {\em volume}-weighted density distribution of the stream, not the mass-weighted distribution. 
To avoid drawing many low-density encounters however, we instead draw from the mass-weighted distribution, but weight each draw by $\mathcal{W}_i = 1/\rho_i$, i.e. $dV/dM$. 
One additional complication of drawing the $\rho_i$ is that because $A_\mathrm{GG} \sim 1$, many draws will have $A_\mathrm{GG} < 1$, which implies that the volume-weighted PDF of $\rho$ diverges at low densities or high volumes. 
We therefore impose a minimum density of $\rho_\mathrm{min} / \rho_0$ of order $10^{-4}$, since the very low-density end of the PDFs is affected by shot noise in the finite sample of ISO tracer particles we use to evaluate the stream density distributions.  

For each population model, we can now construct the rate distribution (Figure~\ref{fig:rate_distr}).
We would like to know the total number of streams, and the total rate $\mathcal{R}$, and how these two quantities are distributed as a function of the rate from each stream $\mathcal{R}_\mathrm{stream}$. 
The cumulative distribution of the number of streams, i.e. the number of streams with an encounter rate greater than $\mathcal{R}_\mathrm{stream}$, is 
\begin{equation}
    N_\mathrm{streams}(\mathcal{R}_\mathrm{stream}) = \frac{1}{N_\mathrm{samples}} \sum_{i=1}^{N_\mathrm{samples}} n_\mathrm{prog} \mathcal{W}_i \mathcal{I}_{\mathcal{R}_i > \mathcal{R}_\mathrm{stream}},
\end{equation}
where $\mathcal{W}_i$ and $\mathcal{R}_i$ are the samples we have drawn, $N_\mathrm{samples}$ is the number of samples we have drawn, and $\mathcal{I}$ is an indicator function which is 1 when the condition in the subscript is true, and 0 when it is false. 
The number density of progenitors, $n_\mathrm{prog}$, converts the weights from a volume to the number of progenitors (and hence streams) contained within such a volume. 
We adopt
\begin{equation}
    n_\mathrm{prog} = \frac{0.1\ M_\odot\ \mathrm{pc}^{-3}}{\langle M_\mathrm{prog} \rangle},
\end{equation}
where the numerator is the density of stellar mass and remnants in the Solar neighborhood \citep{mckee_Stars_2015} adjusted upwards by a factor of 2 to account for the past mass loss of stars over the course of their lifetimes \citep{tinsley_evolution_1980, leitner_fuel_2011, hopkins_Galactic_2023}. 
The denominator is the average mass of the progenitors at the time of their birth in the model being considered. 
Similarly, the cumulative rate, i.e. the contribution to the total rate $\mathcal{R}$ from streams with individual rates greater than $\mathcal{R}_\mathrm{stream}$ is
\begin{equation}
\label{eq:rcumulative}
    \mathcal{R}(\mathcal{R}_\mathrm{stream}) = \frac{1}{N_\mathrm{samples}} \sum_{i=1}^{N_\mathrm{samples}} n_\mathrm{prog} \mathcal{W}_i \mathcal{R}_i \mathcal{I}_{\mathcal{R}_i > \mathcal{R}_\mathrm{stream}}.
\end{equation}

\begin{figure}
    \centering
    \includegraphics[width=\linewidth]{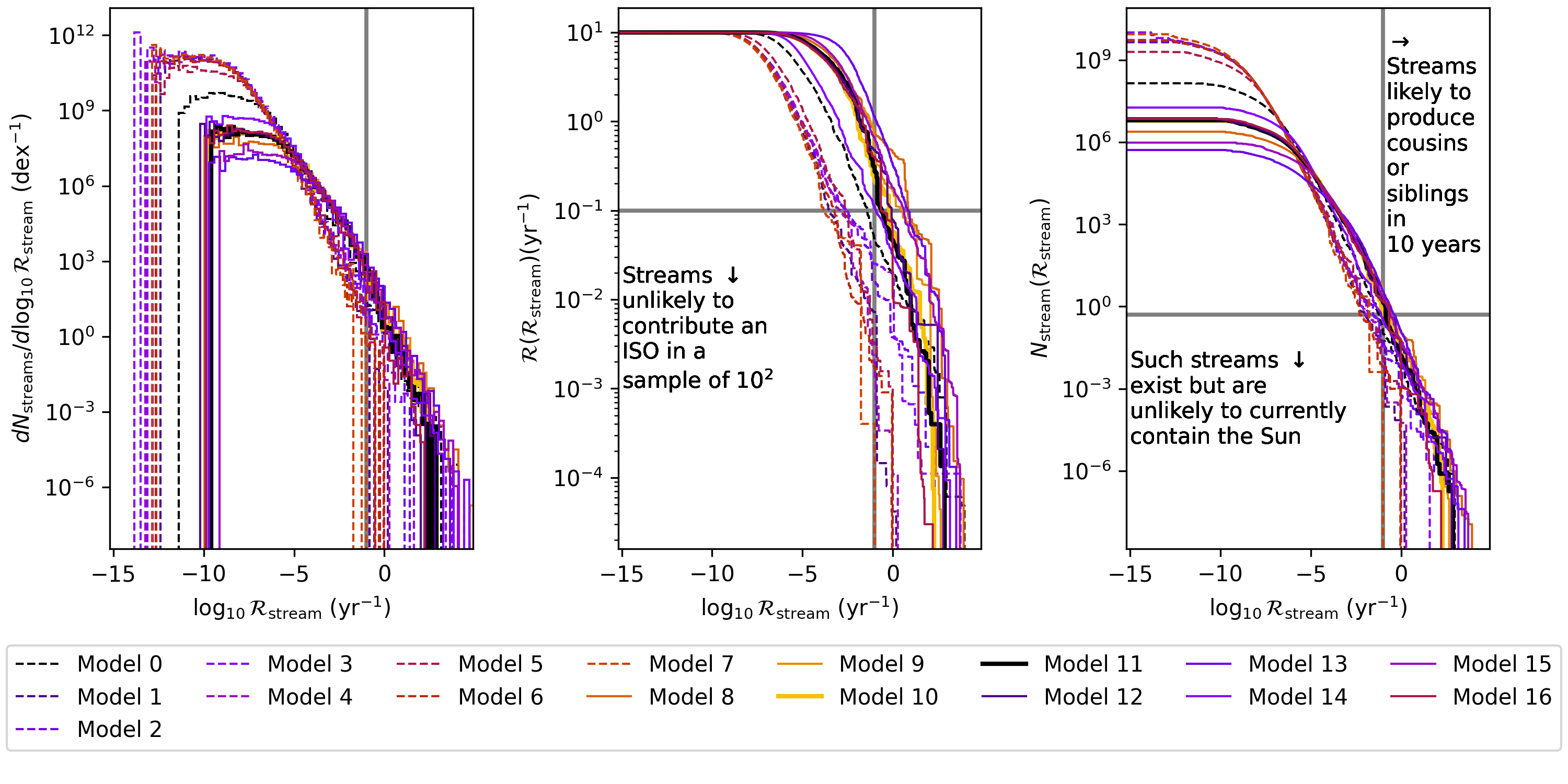}
    \caption{ISO stream rate distributions for normalised-rate $q\le5$~au Solar System encounters over ten years. As a function of the ISO production rate of individual streams, $\mathcal{R}_\mathrm{stream}$, we show the number of streams per dex in $\mathcal{R}_\mathrm{stream}$ (left panel), the cumulative rate $\mathcal{R}$ of ISOs from all streams with rates greater than $\mathcal{R}_\mathrm{stream}$ (middle panel), and the cumulative number of streams with rates greater than $\mathcal{R}_\mathrm{stream}$ (right panel). Each curve represents a different population model listed in Table~\ref{tab:popmodels}, with dashed lines showing models where the streams are produced by individual stars, and solid lines showing models where streams are produced by clusters. To aid in interpretation, we include a vertical line in each plot at $\mathcal{R}_\mathrm{stream} = 0.1\ \mathrm{yr}^{-1}$, which is the rate necessary for any one stream to be reasonably likely to produce multiple observable ISOs in 10 years. We also show a horizontal line at a cumulative rate of $\mathcal{R}(\mathcal{R}_\mathrm{stream})=10^{-2}$ in the center panel, indicating where the cumulative rate has become so low that streams are unlikely to contribute a single ISO in a sample of $10^3$ ISOs. Finally we show where $N_\mathrm{streams}$ drops below $0.5$ in the right panel, indicating that for a given population model, no stream of such a high rate is likely to currently contain the Sun. Models with more of their curves in the upper-right corners of the right two plots, as defined by these vertical and horizontal lines, are more likely to produce multiple ISOs from the same stream.}
    \label{fig:rate_distr}
\end{figure}

\begin{table}
    \centering
    \begin{tabular}{cccccc|cc}
       Model & Prog. Mass & Heating & $\sigma_0$\footnote{The initial velocity dispersion of the stream is drawn from a clipped log-normal distribution with these parameters -- any values below $v_T$ are replaced with $v_T$.} & $\sigma_\mathrm{prog}$ \footnote{The velocity dispersion of the progenitor population $(\mathrm{km}\ \mathrm{s}^{-1})$} & Density Distr.\footnote{def. refers to default values, namely $A_\mathrm{GG} = 1 \pm 0.1$, $C_\mathrm{GG}=1 \pm 0.1$, $B_\mathrm{GG}/\rho_0 = 1 \pm 0.1$, and $\rho_\mathrm{min}/\rho_0=10^{-6}$.} & $\mathcal{N}$\footnote{Adjusted value of $\mathcal{N}$ such that the total rate of ISO encounters for each population is 10 yr$^{-1}$.}$/10^{16}$ & $E(\mathrm{rel.}|10^2\ \mathrm{ISOs})$\footnote{Expected number of progenitor relatives (cousins or siblings) in an ISO population of 100, i.e. after ten years.}\\
       \hline \hline
       0  & Kroupa  & $\mathcal{H}=0.1$ & $v_T$ & 30 & def & 1.53 & 0.59 \\
       1  & Kroupa  & $\mathcal{H}=1$ & $3\ \mathrm{km}\ \mathrm{s}^{-1} \pm 0.3\ \mathrm{dex}$ & 10 & def & 2.45 & 0.044\\
       2  & Kroupa & $\mathcal{H}=1$ & $v_T$ & 30& def & 1.40 & 0.19\\
       3  & Kroupa  & $\mathcal{H}=1$ & $v_T$ & 30 & $\rho_\mathrm{min}/\rho_0=10^{-6}$ & 1.22 & 0.29 \\
       4  & Kroupa & $\mathcal{H}=1$ & \sigdist{3}{0.3} & 30& def & 1.57 & 0.041 \\
       5  & Kroupa & $\mathcal{H}=0.5$ & \sigdist{3}{0.3}  & 30& def & 1.78 & 0.063 \\
       6  &  Kroupa & $\mathcal{H}=1$ & \sigdist{3}{0.3} & 30 & def & 1.70 & 0.027\\
       7  & Kroupa & $\mathcal{H}=1$ & \sigdist{10}{0.3} & 30 & def & 1.71 & 0.025 \\ \hline
       8  & \cl{10^2}{10^6}{-2} & $\mathcal{H}=0.5$ & \sigdist{3}{0.3} & 10 & $B_\mathrm{GG}/\rho_0 = 0.4$ & 2.52 & 7.85 \\ \hline
       9  & \cl{10^2}{10^6}{-2} & $\mathcal{H}=1$ & \sigdist{1}{0.3} &10  & $B_\mathrm{GG}/\rho_0 = 0.4$ & 2.05 & 5.02 \\ \hline
       10  & \cl{10^2}{10^6}{-2} & $\mathcal{H}=1$ & \sigdist{3}{0.3} & 10 & $B_\mathrm{GG}/\rho_0 = 0.4$ & 2.50 & 3.02 \\ \hline
       11  & \cl{10^2}{10^6}{-2} & $\mathcal{H}=1$ & \sigdist{3}{0.3} & 30 & $B_\mathrm{GG}/\rho_0 = 0.4$ & 1.75 & 3.11\\ \hline
       12  & \cl{10^2}{10^6}{-2} & $\mathcal{H}=1$ & \sigdist{3}{0.03} & 10 & $B_\mathrm{GG}/\rho_0 = 0.4$ & 2.34 & 4.64\\ \hline
       13  & \cl{10^2}{10^6}{-1.5} & $\mathcal{H}=1$ & \sigdist{1}{0.3} & 10 & $B_\mathrm{GG}/\rho_0 = 0.4$ & 2.02 & 10.63\\ \hline
       14  & \cl{10^2}{10^6}{-2.5} & $\mathcal{H}=1$ & \sigdist{1}{0.3} & 10 & $B_\mathrm{GG}/\rho_0 = 0.4$ & 2.13 & 1.02\\ \hline
       15  & \cl{10^3}{10^6}{-2} & $\mathcal{H}=1$ & \sigdist{1}{0.3} & 10 & $B_\mathrm{GG}/\rho_0 = 0.4$ & 2.34 & 6.08\\ \hline
       16  & \cl{10^2}{10^6}{-2} & \heat{1}{0.3} & \sigdist{1}{0.3} & 10 & $B_\mathrm{GG}/\rho_0 = 0.4$ & 2.33 & 3.78\\
    \end{tabular}
    \caption{Parameters for the population models of ISO streams.}
    \label{tab:popmodels}
\end{table}

These distributions are shown in the right two panels of Figure \ref{fig:rate_distr}, along with the corresponding PDF of the $N_\mathrm{streams}(\mathcal{R}_\mathrm{stream})$ distribution (left panel). 
The model parameters for each line are shown in Table~\ref{tab:popmodels}.
We remind the reader that each population model is adjusted as follows so that the total rate of ISOs that encounter the Solar System with pericenters below $q_\mathrm{max} = 5$ au is 10 yr$^{-1}$; roughly a rate and observable volume plausible in LSST. 
To adjust the models to meet this constraint, we could in principle alter any of the three factors in Equation \ref{eq:rcumulative}. 
However, $n_\mathrm{prog}$ is well-constrained by observations, and the $\mathcal{W}_i$ follow immediately from $\rho_i$, the clearest quantity to adjust is $\mathcal{N}$, a factor of $\mathcal{R}_i$ --- which represents the number of ISOs produced per solar mass of progenitor. 
Note that we make no dependence on the number density/mass density relationship of ISOs per progenitor \citep{ISSI_2019}; i.e. no ISO size-frequency distributions are implied.
The factor by which $\mathcal{N}$ and hence the overall rate $\mathcal{R}$ is adjusted for each population model is shown in Table~\ref{tab:popmodels}, and is typically about a factor of two.

The cumulative ISO encounter rate and number of streams split up by $\mathcal{R}_\mathrm{stream}$, namely the right two panels of Figure \ref{fig:rate_distr}, are an intermediate step between drawing the sample values of $\mathcal{R}_i$, and deriving the statistics of how frequently a given population model will produce ISO cousins or siblings, which we do below. 
Nonetheless, we can use these distributions to read off which models are likely to produce cousins or siblings that enter our assumed observable sphere (perihelia $\le 5$~au). 
In order to do so, there needs to be a stream that A) has a high enough rate that it can produce multiple ISOs in a reasonable amount of sky surveying time, and B) the stream has a high probability of interacting with the Solar System at any given time. 
These conditions are shown as vertical and horizontal lines respectively in the rightmost panel. 
Populations of streams are more likely to produce cousins or siblings if their distribution includes a component in the upper-right corner of this plot. 
Note that the vertical line is quite conservative, since {\em all} streams to its right are reasonably likely\footnote{The probability is $1-P(0)-P(1)$ where $P(x)$ is the Poisson distribution with expected value $\lambda=1 = (10\ \mathrm{yr})\ \cdot (0.1\ \mathrm{yr}^{-1})$, i.e. $26\%$.} to produce a sibling or cousin in a ten-year survey, whereas many more streams, to the left of our vertical line, each have a (smaller) chance of contributing multiple ISOs.

\begin{figure}
    \centering
    \includegraphics[width=0.9\linewidth]{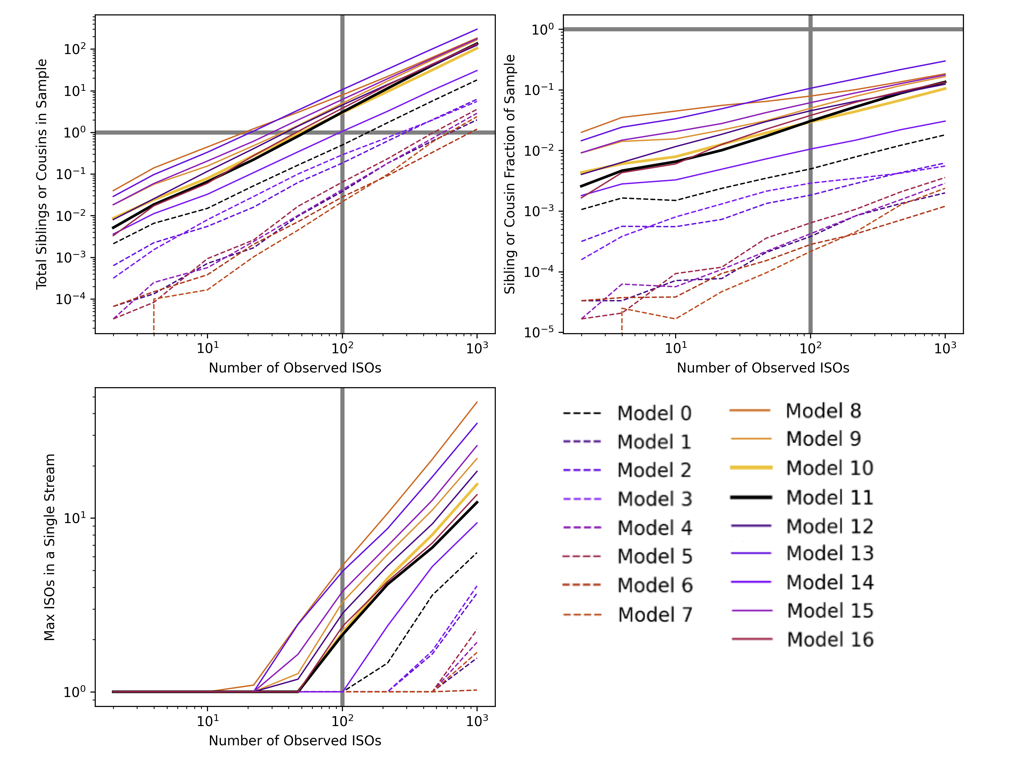}
    \caption{Properties of the ISO `relatives' (cousins or siblings) as a function of the observed sample size of ISOs. The top left panel is the expected number of ISO relatives in the observed population, the top right panel is the fraction of the sample composed of relatives, and the bottom panel shows the maximum number of ISOs from a single stream.    Each line is the average of many simulated sample populations given a population model (see Table \ref{tab:popmodels}). Solid lines show cluster progenitors and dashed lines show single star progenitors. Vertical lines are included at $10^2$, corresponding to 10 years of observations for these populations, all of which are normalized to have a total rate of ISO encounters with the Sun of 10 yr$^{-1}$. We also include horizontal lines at 1 in the first two panels, corresponding respectively to when ISO relatives are likely to be seen and when 100\% of the ISO population has an observed relative.}
    \label{fig:summary_twins}
\end{figure}

We can now use the rate distributions to draw sample populations and determine how common siblings and cousins are. 
A stream with a given $\mathcal{R}_\mathrm{stream}$ has a probability of producing cousins or siblings of 
\begin{equation}
    P(\mathrm{siblings/cousins} | \mathcal{R}_\mathrm{stream}) = 1 - \exp(-\lambda) - \lambda\exp(-\lambda)
\end{equation}
with $\lambda=\mathcal{R}_\mathrm{stream} T_\mathrm{survey}$, and $T_\mathrm{survey}$, the duration of the survey, set equal to $N_\mathrm{ISO}/\mathcal{R}$. 
Therefore in each bin of the rate distribution, we can compute how likely streams in that bin are to produce cousins or siblings. 
For each bin, we separately draw the number of streams in that bin in this instance from a Poisson distribution with expected value $(d N_\mathrm{streams}/d\log_{10}\mathcal{R}_\mathrm{stream})\Delta \log_{10}\mathcal{R}_\mathrm{stream}$, where the first factor is the height of the histogram in the left panel of Figure \ref{fig:rate_distr}, and the second factor is the bin width in dex. 
Now for each bin, we can draw from a binomial distribution with $p=P(\mathrm{siblings/cousins}|\mathcal{R}_\mathrm{stream})$ and $n$ equal to the number of streams in the bin, which we just drew. 
This binomially-distributed variable is the number of streams that produce siblings or cousins in this bin. 
For each bin with such a stream, we can now draw from our sample of rates $\mathcal{R}_i$ in that bin to record the properties of the stream that produced the cousins or siblings.

Counting up the streams with cousins or siblings and their properties, we see a similar dichotomy to that shown in the rate distributions. 
Figure \ref{fig:summary_twins} shows, as a function of the number of observed ISOs, the total number of cousins or siblings in the observed sample, the fraction of the sample made of cousins or siblings, and the maximum number of ISOs from a single stream in the sample. 
Each curve is produced by averaging $5\cdot 10^4$ realizations of the observed population, as described above; so for instance in the top left panel, y-values of 0.1 should be interpreted to mean that that population will produce a sibling or cousin $\approx 10\%$ of the time for the given sample size of ISOs. 
We see that streams produced by clusters have a much higher propensity for producing cousins than stars do for producing siblings, as we expected based on our earlier order-of-magnitude considerations: there are more ISOs in each stream by a factor of the ratio of the average mass of clusters to the average mass of individual stars $\sim10^4 M_\odot/0.4 M_\odot$, depending on the mass functions. 
This effect may be offset if, for example, the cluster streams have higher velocity dispersions on average. 
We also see that the trend with the number of observed ISOs in each plot is basically linear. 
This is to be expected in the limit that cousins or siblings make up a small fraction of the total ISO population. 
Even in the most extreme case (Model 13), cousins only make up $\sim 30\%$ of the ISO sample for a sample of $10^3$ ISOs. 
Generally the fraction of ISO relatives increases with the number of ISOs, since ultimately all ISOs have a large number, $(M_\mathrm{prog}/1\ M_\odot) \mathcal{N}$, of relatives.

\begin{figure}
    \centering
    \includegraphics[width=0.9\linewidth]{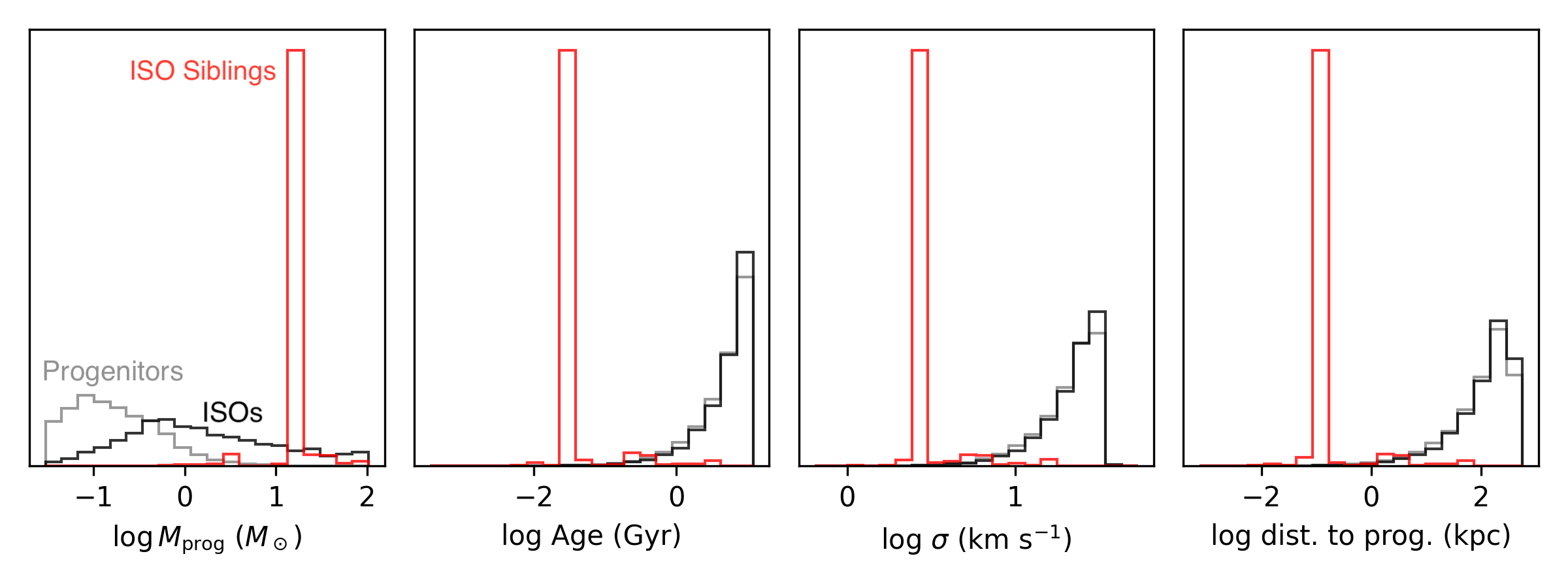}
    \caption{Statistics of encountered `twin' ISOs for single-star progenitors (Model 1 in Table~\ref{tab:popmodels}). Each panel shows the marginal distribution of the given quantity for the progenitors/streams in gray, the ISOs encountered at the Sun in black, and the ISO siblings in red. The left panel shows the log of the progenitor mass, so gray is the IMF, black is the IMF weighted by $M_\mathrm{prog}$, reflecting our assumption that the number of ISOs from a given progenitor is proportional to its mass. The age and velocity dispersion of the stream are directly taken from the model, while the final panel showing distance to the progenitor, is estimated based as the product of the age and velocity dispersion.}
    \label{fig:keyhist_singlestar}
\end{figure}

\begin{figure}
    \centering
    \includegraphics[width=0.9\linewidth]{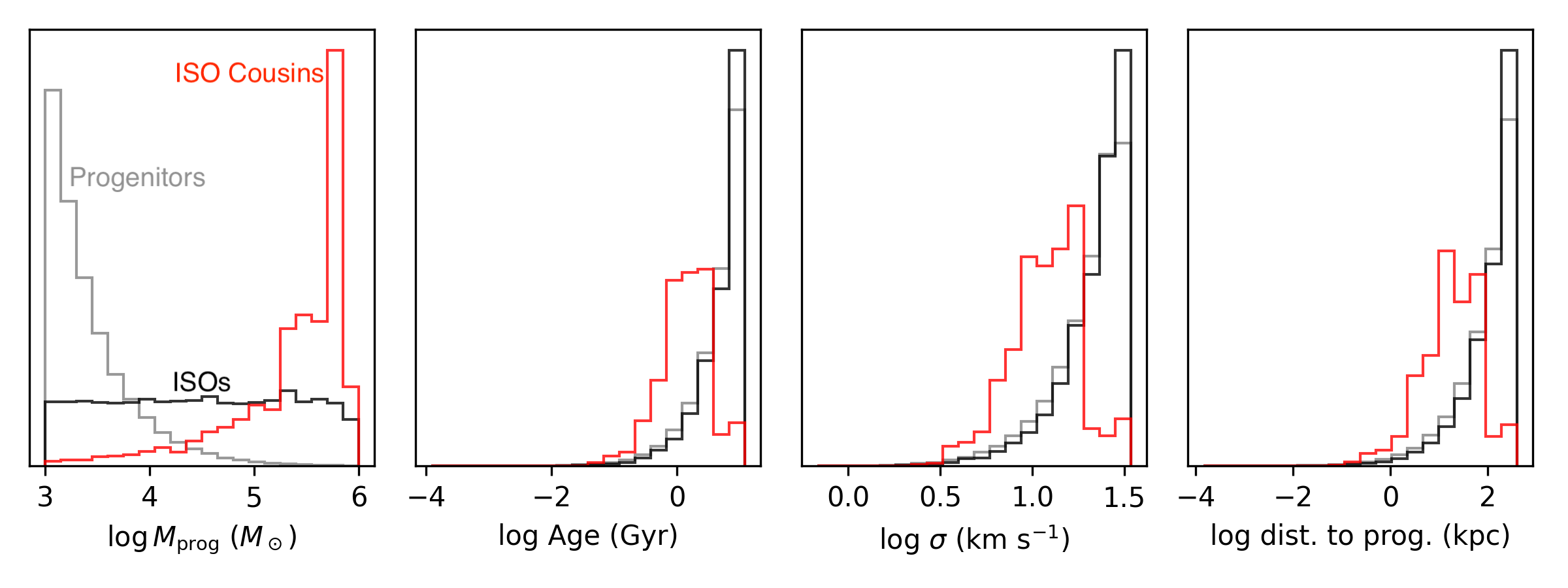}
    \caption{Statistics of encountered `cousin' ISOs for cluster progenitors (Model 15 in Table~\ref{tab:popmodels}). Same as Figure \ref{fig:keyhist_singlestar}, except that the progenitors here are star clusters rather than single stars. }
    \label{fig:keyhist_cluster}
\end{figure}

To investigate which streams end up contributing to the ISO population and the population of cousins and siblings, we can use our simulations of the populations of each to sample from the joint distribution of all stream properties, namely progenitor mass, the time since ejection, the velocity dispersion of the stream, the initial velocity dispersion of the stream, the density parameters, and the LSR velocity of the stream at its intersection with the Sun. 
In Figures~\ref{fig:keyhist_singlestar} and \ref{fig:keyhist_cluster} we show a few key summaries of these distributions for two models: a single star model and a cluster model respectively. 
These histograms show that siblings and cousins are qualitatively quite different. 
Both populations are biased relative to the total ISO population towards the most massive progenitors, though we caution that this is just the result of our earlier assumption that the number of ISOs scales with the mass of the progenitor. 
(It is plausible that the contributed ISOs of a progenitor have an SFD that proceeds nonlinearly with mass; we explore effects on our modelling in \S~\ref{sec:heating}.) 
In the unlikely event that an ISO sibling is present in a sample of 100 ISOs, it must have come from a denser stream, meaning that the age, velocity dispersion, and their product, the typical distance to the progenitor star, must all be relatively low. 
Since in our models a sample of 100 ISOs is highly likely to contain cousins, the cousins end up tracing a similar set of streams to the underlying ISO population, though this bias towards younger and lower-velocity dispersion streams is still present.

We find that in most models the typical distance to the progenitor of an ISO along its stream, namely $\sigma t$, is of order 100 kpc, with a tail extending down to $\sim 1$ kpc. 
Attempts to integrate the trajectory of observed ISOs backwards in time to discover their progenitor system were made for 1I and 2I \citep{gaidos_Origin_2017,portegieszwart_Origin_2018,bailer-jones_Plausible_2018,zuluaga_General_2018,dybczynski_Investigating_2018,zhang_Prospects_2018,bailer-jones_Search_2020,hallatt_Dynamics_2020}. 
Our progenitor distances suggest such approaches will generally not be successful. 
These works attempting to trace 1I/\Ou\ typically acknowledged the difficulty for timescales longer than a few tens of Myr, due to perturbations in the trajectory from GMCs and other non-axisymmetric features of the potential.
One may also interpret our result through the lens that typical ISOs in our population models will have been orbiting the Galaxy for much longer timescales.
However, ISO siblings have much closer typical distances, of order 100 pc, and may be more plausible to retrace.

\section{Discussion}
\label{sec:disc}

The primary observational consequence of ISOs existing in streams rather than as a uniform background is the possibility that we may observe multiple ISOs from the same stream. 
The ISO population of the Galaxy should be viewed as a rich tapestry of overlapping streams that is the outcome of generations of planetesimal formation.
However, to continue our earlier metaphor, drops of water in the mighty river do not trace back to each mountain peak. 
Instead, it is catchment flow and rates that need to be considered, and these are what inform understanding of the time-dependent system evolution.

ISO streams are governed primarily by their internal velocity dispersion $\sigma$. 
In our population models, $\sigma$ is set by a combination of its initial value $\sigma_0$, namely the typical spread in velocities ISOs have when they are ejected from their progenitor system, and the dynamical heating prescription, which increases $\sigma$ over time. 
For reasonable parameter values, both $\sigma_0$ and the heating rate appear to have comparable effects on the statistical distributions of cousins and siblings. 
The single-star model with the greatest chance of producing a sibling within a sample of 100 ISOs is Model 0.
This model has a heating rate $\mathcal{H}=0.1$, a tenth of the value motivated by comparison with the age-velocity dispersion relation, and an initial velocity dispersion $v_T$, which is the minimum plausible value, corresponding to ISOs becoming unbound from the star's Oort cloud due to small perturbations. 
Increasing $\mathcal{H}$ and $\sigma_0$ reduces the incidence of ISO siblings by about a factor of three for each effect, due to the lower prevalence of dense, cold ISO streams. 

\subsection{Ejection Mechanisms}
\label{sec:ejection_mechanisms}

How plausible are the various distributions of $\sigma_0$ and $\mathcal{H}$? First focusing on $\sigma_0$, the typical values we have assumed in many of our population models are $\sim$~1--3~\kms, and in all cases we have assumed that the ISOs are ejected with this velocity immediately upon the star's or cluster's birth. 
These choices rely heavily, but by no means exclusively, on the high-fidelity inferred histories available for the small-body populations of the Solar System.
They are based on dynamical simulations that variously reproduce the properties of the planets, the trans-Neptunian objects (TNOs), and the Oort cloud 
\citep{oort_Structure_1950, kaib_Reassessing_2009}. 
For a recent review of Oort cloud formation and evolution see \citet{kaib_Dynamical_2022}.

The mechanisms by which ISOs may be ejected from their host star system act at different times through a given star's life, so we consider the individual events that can unbind planetesimals, beginning at the point where planetesimals have formed.
Within the first 1-10 Myr, the major event is the dispersal of the gas disk \citep[e.g.][]{pecaut_Star_2016,ansdell_ALMA_2017,pfalzner_Most_2022}. 
Close stellar flybys, largely in the birth cluster, produce asymmetric sprays of unbinding at $\sim 3$~\kms \citep{jilkova_Mass_2016,pfalzner_Significant_2021}, while emplacing orbits akin to extreme TNOs \citep{pfalzner_Trajectory_2024}.
Dynamical instabilities of planetary systems are more energetic; current explorations suggest ISO velocities of 5--10~\kms \citep{adams_Lithopanspermia_2005}. 
Instabilities can be triggered by the gas-disk dispersal \citep{raymond_Origin_2017}, or from resonant perturbations --- in which case, they can most likely be expected within the first few tens of Myr of system life \citep{nesvorny_Dynamical_2018}.
Together, these processes may remove the majority of a system's planetesimals; 
for instance, in one particular model of the Solar System's history, $\sim 90\%$ of the objects ejected from the inner parts of the star system are fully ejected from the system \citep{dones_Oort_2004}.
Some 5\% remain in the Oort cloud; perhaps only $10^{11}$ comets \citep{oort_Structure_1950}.
We can therefore treat the primary source of a stream as occurring as a high rate of supply event: either a single major event, or a sequence of events of similar order of magnitude, spaced closely in time.

An additional factor then comes into play: Galactic tidal unbinding during the main-sequence lifetime will continuously erode a small fraction away from any Oort cloud \citep{Heisler_Tremaine_1986,veras_Great_2014,torres_Galactic_2019}.  
The newly unbound ISOs will plausibly have very low velocities comparable to $v_T$, analogous to stars from clusters at velocities comparable to the internal velocity dispersion of the clusters themselves \citep[e.g.][]{johnston_Prescription_1998,amorisco_Feathers_2015}.
This means that both progenitor situations, individual star systems and clusters, will provide an ongoing low rate-of-supply ``trickle" of dynamically-cold ISOs into their stream\footnote{Unlike braided rivers, neither level of discharge creates any size-frequency distribution sorting of the ISOs.}.
The low rate of supply ``trickle" is however a fractionally small quantity per system, as it comes from the $10^{11}$-strong population retained in each progenitor's Oort cloud, which itself was merely $\sim$1--5\% of the initial planetesimal population \citep{fernandez_Dynamical_1981, dones_Oort_2004, kaib_Formation_2008}.
Regardless of the dynamical heating that a progenitor's initial stream has experienced over time, this continuous low-$\sigma$ contribution will be present --- as long as the system has an Oort cloud. 

The frequency of exo-Oort clouds throughout the Galaxy is little constrained by observation.
Oort cloud presence is theoretically probable, though challenging to confirm observationally \citep{baxter_Probing_2018}; it is strongly suggested by the widespread presence of exoplanetary systems \citep[e.g.][]{fressin_False_2013,dressing_Occurrence_2013}, the abundance of debris disks around stars \citep[e.g.][]{rieke_Decay_2005}, and potentially also the enrichment of white dwarfs \citep{debes_Are_2002}. 
Oort cloud emplacement is expected to occur early.
If the system is embedded in a cluster, this will likely be during the system's residence in its stellar birth cluster \citep[though see][]{portegieszwart_Oort_2021a}.

Episodic higher rate of supply events during Oort cloud erosion will depend on the progenitor's location in the Galaxy; \citet{veras_Great_2014} predict greater erosion from increased flybys near the bulge, while encounters with field stars will strip flyby-dependent fractions \citep{Hanse_2018, pfalzner_Significant_2021}.
Finally, post-main-sequence stellar mass loss will then unbind much of the remainder of the Oort cloud, at velocities of order $v_T$ \citep[e.g.][]{parriott_Number_1998,veras_Great_2011,veras_Solar_2012, Moro-Martin_2019,levine_Interstellar_2023}.

Given the remarkable inefficiency with which Oort clouds form and the high rates of ejection early in each star system's life, the approximation that all ejections occur immediately upon the star's birth appears adequate; but this conclusion may yet be modified upon further refinement of the model to include these effects. 
In particular, by placing a substantial number of ISOs in effectively younger, dynamically colder streams, we might expect more siblings to be seen in future surveys than what we predict here.

Similarly, the early evolution of the streams is more complex than what we have assumed in our population models. By assuming that the peak in the density distribution occurs near $\rho=\rho_0$, we are neglecting the evolution of the stream on timescales between the ejection timescale ($t_f-t_0$ in the language of Section~\ref{sec:sims}) and the timescale for the stream to obtain its full width, $\sim \kappa^{-1}$. During these early times and the timescales where heating has not yet had a chance to increase the velocity dispersion in all directions (see next subsection), the density structure of the stream is sensitive not just to the ejection velocity, but also the time-dependence of ejections and their geometry, i.e. whether ISOs retain a memory of the plane of their birth planetary system, which would reduce the width of the stream in the direction perpendicular to the planetary disk. Again while these effects only apply to a few percent of ISOs, the fact that they act in a direction that makes the streams denser may affect the rate of ISO siblings' and cousins' encounters with the Solar System.

Another model refinement that may end up being important is the unification of the models for the streams from star clusters and from individual stars. 
At present our population models assume that either all streams come from individual stars, or all streams come from clusters. 
Physically these cases correspond to the following two mutually exclusive possibilities:
\begin{itemize}
    \item {\bf Single Stars:} ISOs are ejected from individual stars at velocities well in excess of any cluster potential, or later than the typical cluster lifetime.
    \item {\bf Clusters:} All stars form in clusters and ISOs are ejected early enough and at low enough velocities that they mix in the cluster before they eventually escape and/or the cluster dissolves. Siblings from these stars will be kinematically indistinguishable from their cousins.
\end{itemize}
In reality, neither of these scenarios is likely to be exactly correct. 
Some stars will not form in clusters \citep[e.g.][]{lada_Embedded_2003}, and the ejection mechanism will not be so uniform that all streams will fall into one category or another. 
We defer this unification into a single stream population model to future work.

In any case, the higher values of $\sigma$ from the cluster streams mean that ISOs from the same cluster may arrive in the Solar System with substantially different (of order $\sigma$) velocities and radiants. 
In contrast, ISO siblings will be far closer together in this parameter space, making it plausible that they could be distinguished from chance alignments of unrelated ISOs, at least probabilistically, purely based on their kinematics. 
It may also be possible to chemically distinguish ISOs through their surface composition or coma, such as that seen for 2I/Borisov \citep[e.g.][]{kareta_Carbon_2020, bodewits_Carbon_2020, bannister_Interstellar_2020}.
However, compositional signatures open a broad field of additional complexity. 
With current understanding of disk and planetesimal evolution, it is not yet clear which paths will lead to clean distinctions, given the compositional variation and evolutionary processing within a disk, the variation between disks in a cluster, and the variation between clusters.
For example, ISO cousins may be more- or less-compositionally similar to each other than unrelated ISOs that happen to come from similar regions of their respective protoplanetary disks \citep[e.g.][]{raymond_Survivor_2020,kokotanekova_Life_2023}. 

\subsection{Dynamical Heating}
\label{sec:heating}

We now turn to the question of dynamical heating, which is of similar importance to the initial ejection mechanisms and velocities. 
Our prescription is calibrated such that streams will experience a constant heating rate, that is a constant $d\sigma^2/dt$, for the default values of $\mathcal{H}=1$ and $\alpha_H=0.5$. 
For these values, the magnitude of this heating is such that the oldest stars in the Galaxy, assuming they are subject to the same heating, would reach velocity dispersions of around 35 \kms. This may be compared with the heating rate found by \citet{lacey_influence_1984} in which giant molecular clouds are investigated as the source of the age velocity dispersion relation (AVR) in the Solar neighborhood \citep[e.g.][]{holmberg_genevacopenhagen_2009}. Generally the velocity dispersion of the test particles (stars, or in our case ISOs) increases according to $d\sigma^2/dt = D/\sigma^\gamma$, where $\gamma=0$ corresponds to a constant heating rate, $\gamma=1$ to the case where the scaleheight of the perturbers is greater than the scaleheight of the test particles, and $\gamma=2$ to the more typical case that the perturbers have a smaller scaleheight than the test particles. The velocity dispersion will therefore evolve as $\sigma = ((1+\gamma/2)Dt + \sigma_0^{\gamma+2})^{1/(\gamma+2)}$. The magnitude of the heating for $\gamma=2$ is
\begin{equation} \label{eq:diff}
    D \approx (2/3) G^2 M_c^2 N_c \nu \ln \Lambda \mathcal{F}
\end{equation}
Here $M_c$ is the mass of the perturbers, $N_c$ is the number surface density of perturbers, $\ln\Lambda$ is the Coulomb logarithm, and $\mathcal{F}$ is a correction factor that depends on the local shape of the rotation curve. For a flat rotation curve, $\mathcal{F}\approx 0.47$. For a Solar circle surface density of $\sim 2\ M_\odot\ \mathrm{pc}^{-2}$ \citep{mckee_Stars_2015} divided among a population of giant molecular clouds of typical mass $10^6 M_\odot$, subject to 10 Gyr of heating, with $\ln\Lambda \approx 3$ this heating rate only increases $\sigma$ to about 12 \kms. Moreover, $\gamma=2$ implies $\sigma \propto t^{1/4}$, which is likely lower than the observed value from the AVR. 

Substantial uncertainty in the appropriate values to use in the heating rate \citep{ida_Origin_1993}, and the appropriateness of assuming constant $D$ \citep{aumer_Agevelocity_2016} mean giant molecular clouds may still be an important source of dynamical heating . 
There are also other sources of heating, including perturbations from spiral arms, the bar, and even cold dark matter substructures \citep{bovy_Linear_2017,carlberg_Subhalo_2023}, in rough order of likely importance\footnote{Cold dark matter substructures play a much larger role for observed stellar streams than for ISO streams, because the former are on halo-like orbits and hence not subject to the heating mechanisms at play in the disk.}. 
\citet{aumer_Agevelocity_2016} argue that molecular clouds likely play an important role in increasing the vertical velocity dispersion, whereas spiral arms primarily increase the velocity dispersion only in the plane of the Galaxy, and these two heating mechanisms evolve in their relative importance over the history of the Galaxy (see also \citet{arunima_Their_2025}).
Moreover, they argue that age uncertainties may artificially lower the power law index of $\sigma$'s variation with $t$, as stars at intermediate ages are incorrectly assigned to both older and younger populations. 
On top of uncertainties around the heating rates, it is also possible for older stars to be born with higher velocity dispersions \citep{bournaud_thick_2009, forbes_evolving_2012}, driven by a combination of more intense star formation and feedback and greater degrees of gravitational instability at high redshift \citep[e.g.][]{genzel_sins_2014, krumholz_unified_2018, forbes_How_2023}. 
Given these complications, we stick to a simple empirical relationship that yields a realistically-large velocity dispersion for the default values, leaving the power law scaling with $t$ equivalent to our free parameter $\alpha_H$. 
Reasonable values of $\mathcal{H}$ range from $\sim 0.1$ (corresponding to heating from GMCs only) to $\sim 1$ (the full AVR is explained by heating of some sort), and $\alpha_H$ is likely to be between $0.25$ (GMC heating only) and $0.5$ (constant heating).

An additional question is whether scattering off stars could play a substantial role in dynamical heating. 
In the context of stellar streams or the heating of the stellar disk, this possibility can be dismissed by considering the two-body relaxation time \citep{binney_Galactic_2008}. 
In our case, the masses of the ISOs are instead far smaller than those of the GMCs; so we confirm that stars do not substantially heat ISOs. 
For interactions that can be described by a range of weak distant interactions, we can use Equation \ref{eq:diff} as a rough estimate. 
In the modern Galaxy, the perturber mass of stars is $\sim 10^6$ times smaller than that of GMCs, while the number density of perturbers is $\sim 10^6 \cdot 10^2$ times larger. 
Since $D$ is proportional to $M_c^2 N_c$, the mass effect is far larger. 
While $\ln\Lambda$ is only a few for the GMC case, it is closer to $\approx 17$ in the stellar case, assuming maximum and minimum impact parameters respectively of $b_\mathrm{max} \approx 100\ \mathrm{pc}$ and $b_\mathrm{min} = G (1 M_\odot)/(30\ $\kms$)^2$. 
While this is a large difference, it is still not enough to overcome the large difference in characteristic mass. 
Strong scattering events (which ISOs encountering the Solar System are often experiencing, given that $b_\mathrm{min}\approx 1$~au for the values given above) are exceedingly rare, with any individual ISO having a strong interaction with a star at a rate $\sim n_* |v_\mathrm{rel}| \pi b_\mathrm{min}^2 \sim 10^{-16}\ \mathrm{yr}^{-1}$ --- six orders of magnitude lower than the rate necessary for the typical ISO to experience such an interaction in the history of the Universe. 
A small subset of these ISOs will collide with the star \citep{forbes_Turning_2019}.

An important aspect of the various Galactic heating mechanisms is that they act differently in the vertical and in-plane directions. 
Moreover, the predictions for disk heating are not necessarily directly applicable to streams, because the distribution function of the stream is not the same as that of a stellar disk. 
We have also assumed that we can describe the properties of a stream by a single value of $\sigma$ at each time. 
While this is consistent with our unheated streams from Section \ref{sec:sims}, we have not shown that streams subject to heating can still be described so simply. 
We have also modelled each progenitor as producing exactly $\mathcal{N}(M_\mathrm{prog}/1\ M_\odot)$ ISOs. 
The number of ISOs produced per unit of host stellar mass may, however, be subject to variation \citep{moro-martin_Origin_2018, ISSI_2019}; for example, if some properties of 1I/\Ou\ are explained by tidal disruption events in the progenitor systems \citep{cuk_1I_2018,zhang_Tidal_2020a}. 
In such a case, a progenitor could produce different numbers of ISOs. 
We have experimented with this effect by multiplying each stream's value of $\mathcal{N}$ by a number drawn from a log-normal distribution with scatter $\sigma_\mathcal{N}$, then renormalizing them such that $\mathcal{R}=10\ \mathrm{yr}^{-1}$. 
As expected, increasing $\sigma_\mathcal{N}$ can increase the rate of ISO sibling production. 
For $\sigma_\mathcal{N} \gtrsim 3$ dex, it becomes increasingly common that a single stream accounts for the entire assumed rate of $\mathcal{R}=10\ \mathrm{yr}^{-1}$ --- in which case, every ISO would be a sibling. 
These complications, as well as the potential importance of composing streams of the many different ejection epochs, and incorporating a more realistic progenitor population weighted by the known properties of stars \citep{hopkins_Galactic_2023, hopkins_Predicting_2024} will be the subject of future work.

Besides the occurrence of multiple ISOs from the same stream, another consequence of the inhomogeneity of ISOs is the time-dependence of the encounter rate as the Sun passes through different streams. 
In the ISO context this was explored by \citet{portegieszwart_Oort_2021}, and it also comes up in the context of dark matter mini-halos (\citealt{ohare_Axion_2024}; \citet{dsouza_Enhanced_2024}). 
Given that we have a rate distribution that extends to very large values of $\mathcal{R}_\mathrm{stream}$, it is worth considering how quickly the Sun passes through individual streams and is able to effectively sample the extremes of the rate distribution. 
The Sun will transit through a stream of width $\sigma/\kappa$ on a timescale $\sigma\ \kappa^{-1} v_\mathrm{rel}^{-1}$, where $v_\mathrm{rel}$ is the relative velocity of the Sun and the stream, which will be of order 30 \kms\ for most streams. 
The resulting timescale is $25\ \mathrm{Myr} (\sigma/v_\mathrm{rel})$. 
Typical values of the latter factor are $\sim 1$, though distributional effects are likely to be important, since streams with higher rates also have lower values of $\sigma$. 
The Sun has existed for $\sim 180$ such timescales, so streams with rates up to $\mathcal{R}_\mathrm{stream}$ such that $N_\mathrm{stream}(\mathcal{R}_\mathrm{stream}) \approx 10^{-2}$ are likely to have interacted with the Sun in its history. 
Note that these values are again all relative to our normalisation of the encounter rate of ISOs to 10 yr$^{-1}$.
For many of our population models, this $\mathcal{R}_\mathrm{stream}$ is of order 10 yr$^{-1}$, i.e. a single stream producing of order as many ISOs as the currently-expected rate from all streams. 
This is only a factor-of-two level fluctuation --- so intense ``showers'' of interstellar objects, analogous perhaps in effect but not in cause to the concept of Oort comet showers \citep[e.g.][]{hills_Comet_1981}, are unlikely in Earth's past or future. 
Even for a value of $N_\mathrm{stream}(\mathcal{R}_\mathrm{stream})$ a few orders of magnitude smaller to accommodate distributional effects, the rate distribution is so steep here that $\mathcal{R}_\mathrm{stream}$ appears unlikely to exceed $\sim 100\ \mathrm{yr}^{-1}$.

\section{Conclusions} \label{sec:conclusion}
ISOs from individual stars or star clusters will form streams as they orbit the Galaxy, much as stars from disrupting globular clusters or dwarf galaxies form stellar streams. Here we summarize our main conclusions from this work.
\begin{itemize} 
    \item ISO streams are governed by their internal velocity dispersion. Their density distributions are particularly sensitive to their velocity dispersions, with a peak or characteristic number density of the stream of $(2.5 \sigma)^{-3} \kappa\nu\ t^{-1} (M_\mathrm{prog}/(1 M_\odot)) \mathcal{N}$. In this expression $\sigma$ is the velocity dispersion, $\kappa$ is the local epicyclic frequency, $\nu$ is the local vertical oscillation frequency, $t$ is the age of the stream, $M_\mathrm{prog}$ is the progenitor mass, and $\mathcal{N}$ is the number of ISOs per solar mass of the progenitor.
    \item We combined the density distribution of individual streams, determined in Section \ref{sec:sims}, with empirically-motivated assumptions about dynamical heating in the disk, the number density of stars, and the velocity distribution of ejected ISOs. For each such model of the population of ISO streams, we can compute the expected distribution of ISO encounters with the observable volume around the Sun for each stream. These rate distributions then determine how frequently the Sun is likely to encounter multiple ISOs from the same stream.
    \item The Sun is currently contained in $\sim 10^7$ streams from clusters, many of which may be composed of streams from individual stars, totalling $\gtrsim 10^{10}$ streams. However, these streams contribute wildly different ISO encounter rates to the observable volume, ranging from $10^{-15}$ per year, to $\sim 0.1$ per year. The Sun may occasionally pass through streams with far higher encounter rates.
    \item We find that it is considerably more likely for multiple ISOs in the observed population to have come from the same star cluster (``cousins'') than it is for multiple ISOs to come from the same star (``siblings''). While siblings are likely to have similar incoming velocities and radiants, cousins will be considerably more spread out, due to the higher internal velocity dispersions of their streams, and may be difficult to distinguish from unrelated ISOs.
    \item Observed ISOs that come from the same star will be highly biased towards the densest streams: the stars that produce the largest number of ISOs with the lowest velocity dispersion and subject to the least dynamical heating (the youngest streams).
    \item Almost all ISOs will not be traceable back to their host star, with typical distances to the progenitor (which may no longer be extant) ranging from 1 to several hundred kpc\footnote{Distances are measured along the stream, so distances $\gtrsim 2\pi R_0 \approx 50$ kpc correspond to particles that have wrapped around the Galaxy relative to the progenitor.}. The exception is ISO siblings, whose progenitors may be as close as $\sim 100$ pc.
    \item It is unsurprising that 1I/\Ou\ and 2I/Borisov originated from different parts of the velocity distribution. Our population models expect that samples of $\sim10^2$--$10^3$ detected ISOs will be necessary in order to see the first ISO siblings. The discovery of two ISOs with very similar incoming velocities before this point would place strong and surprising constraints on the kinematic heating history of the Milky Way, the mechanisms by which ISOs are ejected from their parent star systems, and the variation in the ISO production rate per star.
\end{itemize}


M.T.B. and J.C.F. appreciate support by the Rutherford Discovery Fellowships from New Zealand Government funding, administered by the Royal Society Te Ap\={a}rangi. 

R.C.D. acknowledges support from the UC Doctoral Scholarship and Canterbury Scholarship administered by the University of Canterbury, a PhD research scholarship awarded through M.T.B.’s Rutherford Discovery Fellowship, and an LSSTC Enabling Science grant \#2021-31 awarded by LSST Corporation.

M.J.H. acknowledges support from the Science and Technology Facilities Council through grant ST/W507726/1.

We thank Ana Bonaca, Susanne Pfalzner, Chris Gordon, Joe Masiero, Mordecai-Mark Mac Low, Ciaran O'Hare, Hagai Perets, Rosemary Mardling, Garrett Levine, Darryl Seligman, Jay McMahon, Jack Patterson, and Rua Murray for helpful discussions and comments. We also appreciate the thoughtful and constructive report from the anonymous referee.


\software{numpy \citep{vanderwalt_numpy_2011,harris_array_2020},
          matplotlib \citep{hunter_matplotlib_2007},
          scipy \citep{virtanen_scipy_2020},
          }

\appendix

\section{Adaptive Anisotropic Kernel Density Estimator}
\label{app:kde}
In order to characterize the ISO streams, it is useful to be able to estimate their density in both physical and phase space at any particular point in time given a finite number of tracer particles. Our dataset is 6-dimensional, namely the positions and velocities of each tracer particle in cartesian coordinates, and we know that there is substantial anisotropic structure. The dimension of the problem means that even coarse-grained histograms would become prohibitively cumbersome. Kernel density estimators \citep{rosenblatt_Remarks_1956,parzen_Estimation_1962} are a powerful alternative, since no pre-defined grid is necessary. In a generic form, a KDE estimates the probability density as the following sum over the data:
\begin{equation}
\label{eq:kde}
    p(\vec{q}) = \frac1N\sum_i^N K_i( \vec{q} | \vec{q}_i )
\end{equation}
where $K_i(a|b)$ is a probability density function (the Kernel) over $a$, $N$ is the number of points in the dataset, $\vec{q}_i$ is the $i$th element of the dataset, and $\vec{q}$ is an arbitrary point in the same vector space as the data at which we would like to estimate the probability density $p$.

A common choice for $K_i(a|b)$ is a multivariate normal distribution with mean $b$ and covariance equal to a multiple of the identity matrix of the appropriate dimension, with all $K_i$'s identical. The user then only needs to specify the characteristic width of the gaussian, called the bandwidth. This is the approach taken in $\texttt{scipy.stats.gaussian\_kde}$. However, the streams as viewed in 6D phase space (or even 3D space) are of course highly anisotropic. An isotropic kernel would therefore artificially tend either to blur out and thicken features of the stream that are physically narrow, or in an attempt to preserve those features, produce an unphysically patchy density estimate along the stream where the sum in Equation \ref{eq:kde} is dominated by a small number of datapoints.

To alleviate this difficulty, we implement an adaptive kernel density estimator (KDE) where the kernel varies from point to point. We still use a multivariate Gaussian kernel centered on $b$, but we set the covariance matrix of the $i$th point equal to a multiple of the sample covariance $\Sigma_\mathrm{samp}$ of the $N_\mathrm{neigh}$ nearest neighbors:
\begin{equation}
\Sigma_i = \eta_\mathrm{cov} |\Sigma_{i,\mathrm{samp}}|^{\alpha_\mathrm{cov}} \Sigma_{i,\mathrm{samp}}
\end{equation}
The overall scaling $\eta_\mathrm{cov}$ is a constant, and we include another hyperparameter, $\alpha_\mathrm{cov}$ to allow the covariance to scale with the determinant of $\Sigma_{i,\mathrm{samp}}$, i.e. with the local density of points. This is a generalization of the adaptive bandwidth estimator proposed by \citet{abramson_Bandwidth_1982} and \citet{abramson_Arbitrariness_1982} for which $\alpha_\mathrm{cov}=-1/2$ \citep[see][for a general discussion of multivariate adaptive bandwidths]{sain_Multivariate_2002}.
This choice increases the importance of the scaling of each parameter, since now the distance between points not only matters in the evaluation of $K_i$, but also in determining exactly which points are the $N_\mathrm{neigh}$ nearest neighbors, and hence the covariance matrix of each $K_i$. We also need to specify the value of $N_\mathrm{neigh}$, with smaller numbers being more adaptive to the local density, but also more subject to randomness in the tracers. Regardless, we build a kd tree using $\texttt{scipy.spatial.KDTree}$ \citep{Maneewongvatana_Its_1999} when the KDE is initialized to efficiently find the nearest neighbors of each point.

To determine the nearest neighbors of each point, we need to define a distance metric. Since each dimension in the problem may behave differently (and indeed the velocities and positions are not even dimensionally compatible), we first need to scale, at the very least, the velocities. By the axisymmetry of the assumed potential, $x$ and $y$ should scale together, as should $v_x$ and $v_y$. To scale everything to be compatible with $x$ and $y$, we only need 3 additional (hyper)parameters. We denote the scaled version of each quantity $\tilde{x}=x/f_x$, $\tilde{y}=y/f_y$, etc., and we set $f_x=f_y=1$ and $f_{vx}=f_{vy}$. The distance metric used to determine nearest neighbors is then just taken to be the Euclidean distance in the transformed space, $\sqrt{(\Delta\tilde{x})^2 + (\Delta\tilde{y})^2 + (\Delta\tilde{z})^2 + (\Delta\tilde{v}_x)^2 + (\Delta\tilde{v}_y)^2 + (\Delta\tilde{v}_z)^2}$. Internally the KDE works in this transformed space, so the probability density of any point scales as follows:
\begin{equation}
\label{eq:kdescaled}
    p(\vec{q}) =\left(\prod_d f_d \right)^{-1} \frac1N \sum_{i}^N K_i( \vec{\tilde{q}} | \vec{\tilde{q}}_i ).
\end{equation}
Here $d$ indexes the dimension, and $f_d$ is the factor by which the $d$th coordinate is divided when scaling into the KDE's internal coordinate system $q \rightarrow \tilde{q}$.

We set the values of the KDE's hyperparameters ($N_\mathrm{neigh}$, $\alpha_\mathrm{cov}$, and $\eta_\mathrm{cov}$) via 5-fold cross-validation. In principle the scaling of each dimension could also be set via cross-validation, but that dramatically increases the dimensionality of the problem. We therefore instead choose physically-motivated scales and only cross-validate 2 hyperparameters. In particular, we set $f_x=f_y=1$, $f_z=\nu(r_0)/\kappa(r_0)$, $f_{vx}=f_{vy}=\kappa(r_0)$, and $f_{vz}=\nu(r_0)$, where $r_0$ is a representative galactocentric radius of the stream, which we set to 8.1 kpc for streams near the Solar neighborhood. To carry out the cross-validation, the dataset is divided at random into $N_k=5$ equally-sized subsets. Each of these is, in turn, excluded from the remaining $N_k-1=4$ subsets. The 4 subsets are merged into a single training set, which is used to produce a KDE via Equation \ref{eq:kde}. The excluded subset is then used to test this KDE - its cross-validation error is estimated following \citet{sain_Bias_2001} via
\begin{equation}
\label{eq:UCV}
    \mathrm{UCV} = \frac1{N_k} \sum_{k=1}^{N_k} \int p_{-k}(q)^2 d\vec{x}d\vec{v} - \frac2{N_k} \sum_{k=1}^{N_k} \frac{1}{N_{i_k}}\sum_{i_k} p_{-k}(q_{i_k}).
\end{equation}
We refer to the KDE trained on all but the $k$th fold of the cross-validation as $p_{-k}$. Within the $k$th fold, each of the $N_{i_k}$ data points is indexed by $i_k$. The UCV is derived from the requirement that the hyperparameters should minimize the mean squared error of the KDE, namely
\begin{equation}
    \mathrm{MSE} = \int (p(\vec{q}) - \tilde{p}(\vec{q}))^2 d\vec{x} d\vec{v},
 \end{equation}
 where $\tilde{p}(q)$ denotes the true underlying probability distribution we are trying to approximate with the KDE. The true distribution is not known. Expanding the integrand, the term that depends only on $\tilde{p}(q)$ can be neglected because it is not affected by our choice of KDE. The first term is retained in the definition of the UCV, and the second term $-2p(\vec{q})\tilde{p}(\vec{q})$ is asymptotically approximated by the second term in the UCV in the limit of large $N_k$ and overall number of samples.
 
Intuitively the second term of the UCV evaluates the performance of the KDE on the points not included in the training of $p_{-k}$ -- the larger this term, the less ``surprised'' the KDE is to find an example data point at $q_{i_k}$. The first term evaluates the clumpiness of the KDE, with larger values exhibiting more extreme variability from point to point over the entire phase space volume. For each $k$, the integral in the first term can be evaluated efficiently via importance sampling \citep[e.g.][]{newman_Monte_1999},
\begin{equation}
    \int p_{-k}(q)^2 d\vec{x}d\vec{v} \approx \frac{1}{N_\mathrm{MC}}\sum_{i=1}^{N_\mathrm{MC}} p_{-k} (q_i) 
\end{equation}
where the $q_i$ are sampled from the KDE $p_{-k}$, and $N_\mathrm{MC}$ is chosen to be large enough that the estimated error is $\la$ a few percent of the UCV for all hyperparameter variations. The required $N_\mathrm{MC}$ turns out to be $\sim 1000$ typically.

The expression for UCV (equation \ref{eq:UCV}) is minimized with respect to $N_\mathrm{neigh}$ and $\eta_\mathrm{cov}$. Optimizing over the former is much more expensive because a new (though overlapping) set of neighbors needs to be found, and the covariance matrices at each point re-computed. Searching through different values of $\eta_\mathrm{cov}$ is much easier, since the sum in Equation \ref{eq:kdescaled} can quickly be re-evaluated with an alternative value of $\eta_\mathrm{cov}$. 


We also implement the ability to resample from the KDE, i.e. draw an arbitrary number of new locations in 6D phase space sampled from the probability density. For each desired new sample, we choose one of the original datapoints at random with equal probability, then sample from that point's normal distribution. The newly-drawn point is then rescaled to transform from the KDE's internal scaled coordinates back to the data's original unscaled coordinates.

In addition to estimating the phase space density, that is the fraction of tracers per unit volume, per unit velocity$^3$, also called the distribution function $f(\vec{x},\vec{v})$, it is also useful to both estimate and draw random samples from the distributions 
\begin{equation}
    p(\vec{x}) = \int f(\vec{x}, \vec{v}) d\vec{v},
\end{equation}
namely the marginal distribution, i.e. the density in physical space, and the conditional distribution:
\begin{equation}
    p(\vec{v}|\vec{x}) = f(\vec{x}, \vec{v})/p(\vec{x}).
\end{equation}
The latter is the distribution of velocities at any given location in physical space. To understand these distributions, it will be helpful to split the covariance matrix at each point, $\Sigma_i$ into the following four matrices
\begin{equation}
    \Sigma_i = \begin{pmatrix}
        \Sigma_{i,xx} & \Sigma_{i,xv} \\
        \Sigma_{i,vx} & \Sigma_{i,vv}
    \end{pmatrix},
\end{equation}
each representing a 3 by 3 subset of $\Sigma_i$. The subscript $x$ denotes the first 3 rows or columns, namely the part of $\Sigma_i$ encoding the distribution in physical space, while the subscript $v$ denotes the last 3 rows or columns, the part that deals with the three components of the velocity.

The marginal distribution is then just
\begin{equation}
    p(\vec{x}) = \frac1N \sum_i \mathcal{N}(\vec{x} | \vec{x}_i, \Sigma_{i,xx} ),
\end{equation}
where $\mathcal{N}(x|a,B)$ denotes the probability density function over $x$ of a multivariate normal distribution with mean $a$ and covariance $B$.

The conditional distribution's PDF $p(\vec{v}|\vec{x})$ is straightforward to evaluate from its definition, since we already have the means to estimate both $f(\vec{x},\vec{v})$ and $p(\vec{x})$. However, in order to draw samples from $p(\vec{v}|\vec{x})$ it is useful to compute the properties of this distribution explicitly. We use the standard result that the conditional distribution of a subset of a multivariate gaussian, conditioned on the other elements of the multivariate gaussian, is itself a gaussian. In this case,
\begin{equation}
    p(\vec{v}|\vec{x}) = \frac1N\sum_i \mathcal{N}(\vec{v}|\vec{v}_i + \Sigma_{i,vx} \Sigma_{i,xx}^{-1} (\vec{x} - \vec{x_i}), \Sigma_{i,vv} - \Sigma_{i,vx}\Sigma_{i,xx}^{-1} \Sigma_{i,xv} )
\end{equation}
By precomputing these modified covariance matrices for each point at the time of initialization, we can quickly draw random samples from this distribution, which is useful for monte carlo evaluation of integrals over the velocity distribution.

\section{Convergence of the density distribution}
\label{app:convergence}
\begin{figure}
    \centering
    \includegraphics[width=0.6\linewidth]{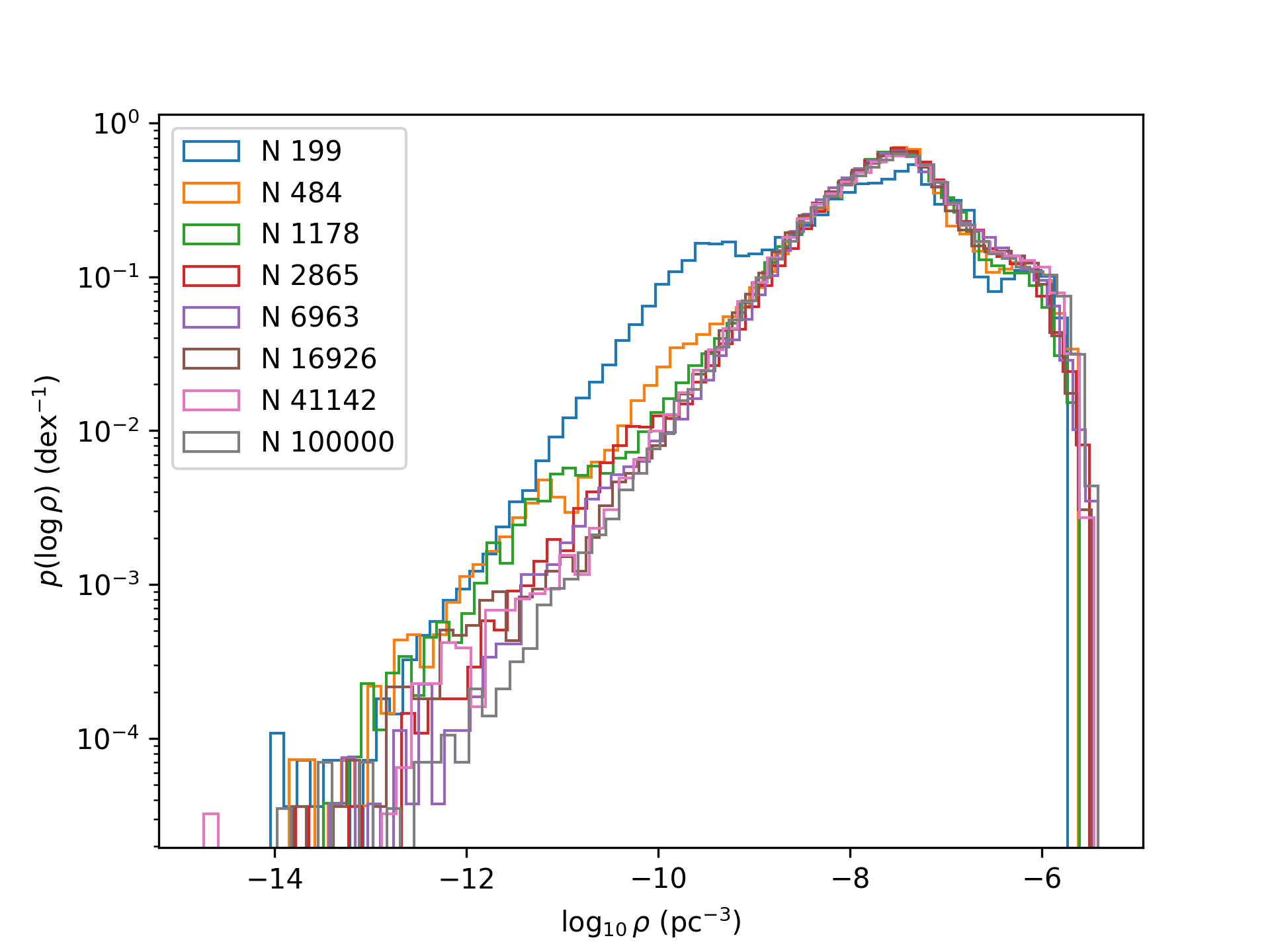}
    \caption{Convergence of the KDE. Each color shows the density distribution estimated from an adaptive KDE trained on an increasing number of samples.}
    \label{fig:density_convergence}
\end{figure}

The density distribution of an individual ISO stream is of central importance in this work. We estimate this distribution by constructing a 6D adaptive kernel density estimator. Given that the cross-validation statistic that we maximize to fix the hyperparameters of the KDE is itself proportional to an estimator of the mean squared error difference between the KDE and the true density distribution, we have every expectation that it will behave well as the number of sample points $N$ increases. Nonetheless, we check explicitly for the density distribution at a particular time for a particular stream. This is the $\sigma=1\ \mathrm{km}\ \mathrm{s}^{-1}$ stream with $M_\mathrm{prog}=1\ M_\odot$ at $t=4\ \mathrm{Gyr}$. We show the density distribution for logarithmically-spaced numbers of points $N$ in Figure \ref{fig:density_convergence}. Each sample is chosen randomly from the $10^5$ particles in the stream, drawn without replacement. That is, the $N=6963$ particles drawn for the purple line have no duplicates, but they will have particles in common with the other samples (in fact they must since the $N=10^5$ sample contains all the particles).

The figure shows that the distribution converges reasonably well for $N\gtrsim 3000$ particles, especially away from the low-density tail. We can also see that for this particular time and this particular stream, the generalized gamma distribution is a poor fit to the density distribution, which is not quite bimodal, but seems to contain another high-density component. Perhaps because of this, there is substantial scatter in the best-fit values of the generalized gamma distribution as a function of $N$ in this case, limiting our ability to estimate an explicit rate of convergence of these quantities.

\newpage
\bibliography{AddRefsHere,Epicycles}
\bibliographystyle{aasjournal}

\end{document}